\documentclass[prd,amsmath,superscriptaddress,twocolumn,nofootinbib]{revtex4-2}
\pdfoutput=1
\usepackage[utf8]{inputenc}
\usepackage{graphicx}
\usepackage{float}
\usepackage{epsf,latexsym,bbm,euscript}
\usepackage{amssymb,amsmath}
\usepackage{mathtools} 
\usepackage{times}
\usepackage{soul,xcolor}
\usepackage{mathtools}
\usepackage{mathrsfs}
\usepackage{array}
\usepackage{booktabs}
\usepackage{enumitem}

\usepackage[caption=false]{subfig}
\newcommand{\subfigimg}[3][,]{%
	\setbox1=\hbox{\includegraphics[#1]{#3}}%
	\leavevmode\rlap{\usebox1}%
	\rlap{\hspace*{-18pt}\raisebox{.5\baselineskip}{\small{#2}}}%
	\phantom{\usebox1}%
}

\newcommand{\RN}[1]{\uppercase\expandafter{\romannumeral #1\relax}}

\usepackage{environ}
\NewEnviron{thincases}{\scalebox{0.5}[1]{$\displaystyle
		\left\{\scalebox{2}[1]{\setlength{\arraycolsep}{6pt}
			$\displaystyle\begin{array}{ll}
				\BODY
			\end{array}$}\right.$}}
			\NewEnviron{thinpmatrix}{\scalebox{0.5}[1]{$\displaystyle
	\left(\scalebox{2}[1]{$\displaystyle
		\begin{matrix}
			\BODY
		\end{matrix}$}
	\right)$}}
	
	\makeatletter
	\g@addto@macro\bfseries{\boldmath}
	\makeatother
	
	\makeatletter
	\newcommand*{\defeq}{\mathrel{\rlap{%
		\raisebox{0.3ex}{$\m@th\cdot$}}%
	\raisebox{-0.3ex}{$\m@th\cdot$}}%
=}
\newcommand*{\eqdef}{=\mathrel{\rlap{%
		\raisebox{0.3ex}{$\m@th\cdot$}}%
	\raisebox{-0.3ex}{$\m@th\cdot$}}%
	}
	\makeatother

	\usepackage{pifont}

	\usepackage{scalerel}
	\usepackage{tikz}
	\usetikzlibrary{svg.path}
	\definecolor{orcidlogocol}{HTML}{A6CE39}
	\tikzset{
orcidlogo/.pic={
	\fill[orcidlogocol] svg{M256,128c0,70.7-57.3,128-128,128C57.3,256,0,198.7,0,128C0,57.3,57.3,0,128,0C198.7,0,256,57.3,256,128z};
	\fill[white] svg{M86.3,186.2H70.9V79.1h15.4v48.4V186.2z}
	svg{M108.9,79.1h41.6c39.6,0,57,28.3,57,53.6c0,27.5-21.5,53.6-56.8,53.6h-41.8V79.1z M124.3,172.4h24.5c34.9,0,42.9-26.5,42.9-39.7c0-21.5-13.7-39.7-43.7-39.7h-23.7V172.4z}
	svg{M88.7,56.8c0,5.5-4.5,10.1-10.1,10.1c-5.6,0-10.1-4.6-10.1-10.1c0-5.6,4.5-10.1,10.1-10.1C84.2,46.7,88.7,51.3,88.7,56.8z};
}
}
\newcommand\orcidlink[1]{\href{https://orcid.org/#1}{\mbox{\scalerel*{
			\begin{tikzpicture}[yscale=-1,transform shape]
				\pic{orcidlogo};
			\end{tikzpicture}
		}{X}}}}
		
		\usepackage{url,hyperref}
		\hypersetup{colorlinks,linkcolor={blue!55!black},citecolor={red!50!black},urlcolor={blue!45!black},breaklinks=true}

\def\cO{\mathcal{O}}

\usepackage{url,hyperref}
\hypersetup{colorlinks,linkcolor={blue!55!black},citecolor={red!50!black},urlcolor={blue!45!black}}

\begin{document}

	\title{Euclidean methods and phase transitions for the strongest deformations compatible with Schwarzschild asymptotics}
	
	\author{Ioannis Soranidis\orcidlink{0000-0002-8652-9874}}
	\email{ioannis.soranidis@hdr.mq.edu.au}
	
	\affiliation{School of Mathematical and Physical Sciences, Macquarie University, Sydney, New South Wales 2109, Australia}
	
	\begin{abstract}
		In this paper, we investigate the thermodynamic properties of a regular black hole model which exhibits the most significant subleading corrections to the Schwarzchild asymptotic behavior, in the context of general relativity, using the Euclidean path integral approach. We review the derivation of the Lagrangian for the matter fields which act as a source for this geometry, explicitly derive the proper thermodynamic quantities introduced in the first law of black hole mechanics, and show that they satisfy the Smarr formula. This analysis naturally leads to the emergence of an effective temperature that is distinct from the one associated with surface gravity. Furthermore, we study the phase structure in anti-de Sitter, Minkowski, and de Sitter spacetimes in the canonical ensemble, considering this effective temperature as the appropriate choice. We show that in this case the regularization of the singularity prevents the Hawking-Page transition and also leads to a deviation from the ``universal" mean-field theory critical ratio. We conjecture that the way a singularity is rendered smooth plays a pivotal role to the degree of this deviation. Finally, we provide remarks on constraints imposed on the minimal length scale by observational data and the viability of regular black holes.
	\end{abstract}
	\maketitle
	
	\section{Introduction}\label{sec:introduction}
	
	Despite the incontrovertible evidence for the existence of dark massive ultracompact objects \cite{GMBTK:00, Sall:02, LIGO:21, B:17, EHT:19, IM:19, RN:03}, their true identity remains elusive, with a plethora of scenarios under consideration \cite{CP:19, BCNS:19, M:23}. The most commonly used type of black holes for comparison with available observational data are mathematical black holes, which are solutions of the Einstein equations in general relativity. Their hallmark features, which make them distinct from other categories like regular black holes (RBHs) \cite{B:68, D:92, H:06}, gravastars \cite{MM:04,MM:23}, wormholes \cite{E:73,MT:88,SV:19}, and fuzzballs \cite{LM:02, M:05} are the existence of an event horizon and a singularity. The problems associated with these two characteristics are the teleological nature of the event horizon and breakdown of general relativity at the singularity.   
	
	The former of these issues can be dealt with by replacing event horizons with quasilocally observable apparent horizons \cite{V:14}, while the latter one is resolved in two possible ways, which arise from the requirement of violation of Penrose's singularity theorem or more specifically at least one of its assumptions \cite{P:65}. The singularity is inherently linked to the geodesic incompleteness of the geometry, so if we want to successfully eliminate it, we need to sufficiently modify the geometry surrounding the focusing point, i.e.,~the singularity. This can be achieved by two methods, namely either with the creation of a defocusing point at finite or infinite affine distance, or by displacing the focusing point at infinity \cite{CRFLV:23}. The first case corresponds to regular black holes with the defocusing point coinciding with its inner horizon in spherically symmetric cases.
		
	 We assume the existence of a full quantum gravity theory that can lead to such a regularization, but its quantum nature is restricted to a finite domain, possibly of Planckian scale. The outcome of such a theory should be a globally hyperbolic and regular geometry, but since we do not possess such a theory yet, we turn our focus on how to source geometries of this nature in the context of general relativity. Historically, nonlinear electrodynamics (NED) was first used by Born and Infeld \cite{BI:34} to cure the infinities/singularities associated with the self-energy of a point charge, but in a large number of cases these types of theories coupled to general relativity are sufficient to generate regular geometries and eliminate singularities only at the cost of using exclusively magnetic charge \cite{B:01}. The first source for such a geometry was found in Ref.~\cite{ABG:00} for the Bardeen black hole and further generalized to a variety of two-parameter families of spherically symmetric RBHs solutions in Ref.~\cite{FW:16}. One notable issue with this type of sources is that magnetic monopoles have not been observed in nature despite significant observational efforts \cite{themacrocollaboration2002,balestra2008,AAB:22}. Additionally, some of these theories fail to provide a Maxwell weak-field limit \cite{B:01}, but exceptions do exist, and the model under consideration in this paper falls into this category. 
	 
	 In the case of classical singularities associated with the point charge, quantum electrodynamics was sufficient to remove them but at the same time make new predictions, which is a key feature of a successful theory, and therefore replaced the NED theories. Similar behavior is anticipated in the gravitational theory with general relativity to be replaced by a more fundamental theory, namely a full quantum gravity theory. In the absence of such a theory, NED coupled to general relativity may prove a valuable model/tool to extract features of quantum gravity. Based on this assumption, the magnetic charge will not be considered a fundamental parameter in our analysis, but rather the minimal length scale will be treated as one. By taking this approach, we can gain valuable insights into the quantum gravity theory within the framework of general relativity.
	 	 
	 The four laws of black hole mechanics were first derived in Ref.~\cite{BCH:73} and close analogies with the four laws of thermodynamics were established. This important link connecting the two fields has since proven to be an important tool in advancing our understanding of black holes. In particular, the physical insights revealed in the rigorous mathematical derivation of the first law in the integral and differential formalism of Ref.~\cite{BCH:73} have provided strong motivation for further investigating their thermodynamic properties. If we believe in a quantum gravity theory which smooths out the singularity, the study of thermodynamic properties of RBHs cannot be an exception. To have a complete understanding of RBHs' thermodynamic properties necessitates the study of a plethora of these models which differ in the way they achieve singularity regularization and extract as much information as possible in order to establish some kind of universal behavior of the singularity smoothing and how it propagates into the classical sector.

	 In this paper, we will study a certain geometry proposed in Ref.~\cite{CLMMOS:23}. The significance of investigating its properties lies in two distinct attributes that set it apart from RBH models proposed by Bardeen and Hayward: Firstly, the NED theory employed to generate this geometry admits the Maxwell weak-field limit, which is a highly desirable property in this type of theories. Potential corrections to the well-established Maxwell theory may manifest in higher energy regimes, while remaining absent in lower energies, giving rise to the conventional Maxwell theory. It is worth pointing out this is not the case for Bardeen or Hayward model which admit a weak-field limit stronger than Maxwell's theory. Therefore, this model is a better candidate for the description of actual astrophysical black holes, since it admits a weak-field limit compatible with a well-established theory. Secondly, this model has the strongest subleading corrections to the Schwarzschild asymptotic behavior, and therefore surpass the ones exhibited by Bardeen or Hayward RBHs. This is a noteworthy feature because effects of singularity smoothing will be more pronounced, making comparison with real astronomical data feasible. Based on these two features we try to establish to what extent there is a deviation of this model from the Bardeen model, studied in Appendix \ref{sec:app:Bardeen}, and the Hayward model studied extensively in Ref.~\cite{SS:23}. Since in a number of black hole cases in anti-de Sitter (AdS) we have seen remarkable equivalence with the liquid-gas critical behavior, we mainly focus on the study of the mean-field theory critical ratio and exponents.

	 The paper is organized as follows: in Sec.~\ref{sec:S2 metric} we describe how to generate the geometry using NED coupled to general relativity embedded in a spacetime with a cosmological constant by explicitly calculating the Lagrangian density\footnote{To avoid confusion, we explicitly mention that we use the terms ``Lagrangian" and ``Lagrangian density" interchangeably throughout this article, but we always mean the Lagrangian density. }. In Sec.~\ref{sec:thermodynamics} we perform a detailed analysis of the Euclidean action and how to derive the proper thermodynamic quantities. In the beginning of Sec.~\ref{sec:1st-law}, we provide some arguments on the reason we allow the cosmological constant and minimal length to vary, and then move on to the study of phase transitions. In Sec.~\ref{sec:PT-AdS}, we study the phase structure and thermodynamic stability of this model embedded in anti de Sitter spacetime along with its behavior near the critical point. In Sec.~\ref{sec:PT-M}, we analyze the phase structure in asymptotically Minkowski spacetime with the implementation of an isothermal cavity to establish thermodynamic equilibrium. We also comment on the bounds imposed on the minimal length and its comparison with astronomical data, leading us to conclusions about the viability of RBHs. We perform a similar thermodynamic analysis in Sec.~\ref{sec:PT-dS} for embedding in de Sitter spacetime with the isothermal cavity restricted between the outer black hole horizon radius and the cosmological horizon.  In all of the above cases we demonstrate the existence of a proper Smarr formula \cite{S:73} and first law of black hole thermodynamics. We conclude in Sec.~\ref{sec:conlusions} with a discussion of our main results and their ramifications, along with potential future research directions. Throughout this article, we restrict ourselves to four spacetime dimensions and we work in dimensionless units such that $\hbar=c=G=k_{B}=1$.

	\section{Regular black hole model}\label{sec:S2 metric}
	In this paper, we focus on the spherically symmetric RBH model proposed in Ref.~\cite{CLMMOS:23}, which is described by a line element given by 
	\begin{align}
		ds^2=-f(r)dt^2+\frac{dr^2}{f(r)}+r^2 d\Omega_{2}, \label{eq:S2:metric}
	\end{align}
	where $d\Omega_{2}=d\theta^2+\sin^2{\theta}d\phi^2$ and 
	\begin{align}
		f(r)=1-\frac{2mr^2}{(r+l)^3}-\frac{\Lambda}{3}r^2,\label{eq:f(r)}
	\end{align}
	with $\Lambda$ being the cosmological constant and $l$ the minimal length scale with the Schwarzschild limit attained for vanishing minimal length scale, i.e.,~when $l=0$. We now explicitly derive the matter Lagrangian that serves as a source for this geometry. It is established that the majority of RBH models are generated by NED coupled to gravity \cite{FW:16, Bambi:book:23,B:01} with the inclusion of magnetic charge. Nevertheless, it is worth noting that certain RBH metrics demand the presence of additional fields beyond the NED theory. An illustrative example of such a case is the Simpson-Visser metric \cite{SV:19}, which interpolates between an RBH and a wormhole, where an additional phantom scalar field \cite{BW:22,RS:23} is present in the matter content. For the case under consideration, we have the following action  
		\begin{align}
		I=\frac{1}{16\pi}\int d^4x\sqrt{-g}(R-\mathcal{L}(\mathcal{F},\Lambda)),
	\end{align}
	with $R$ being the Ricci scalar and $\mathcal{L}(\mathcal{F},\Lambda)$ the Lagrangian density for the matter content which is comprised of the NED theory along with the cosmological constant component. For consistency, we require  this Lagrangian to reduce to $2\Lambda$, when the minimal length scale vanishes. Therefore, we have 
	\begin{align}
		\lim_{l\rightarrow 0}\mathcal{L}=2\Lambda, 
	\end{align}
	which is equivalent to the NED theory being absent.  This Lagrangian is in general some NED Lagrangian, which is a function of the field strength $\mathcal{F}=F^{\mu\nu}F_{\mu\nu}$ combined with the cosmological constant term. The electoromagnetic field tensor is defined in the usual way from the relation
	\begin{align}
		F_{\mu\nu}=\partial_{\mu}A_{\nu}-\partial_{\nu}A_{\mu}.
	\end{align}
	In our analysis, we examine a metric that exhibits spherical symmetry. The commonly adopted form of the vector potential $A_{\mu}$ in this case is given by 
	\begin{align}
		A_{\mu}=\left(-\phi(r),0,0,Q_{m}\cos{\theta}\right),
	\end{align} 
	which corresponds to the presence of an electric and a magnetic charge \cite{FW:16}. To generate this geometry, we only need the presence of a magnetic charge $Q_{m}$, and thus the electric potential $\phi(r)$ will vanish. Therefore, we are led to the potential 
	\begin{align}
		A_{\mu}=\left(0,0,0,Q_m\cos{\theta}\right).\label{eq:Am}
	\end{align}
	The only nonvanishing components of the electromagnetic strength tensor are
	\begin{align}
		F_{23}=-F_{32}=-Q_{m}\sin{\theta},
	\end{align}
	which leads to the field strength of the following form
	\begin{align}
		\mathcal{F}=\frac{2Q_m^2}{r^4}. \label{eq:F(r)}
	\end{align}
	To determine the appropriate Lagrangian $\mathcal{L}(\mathcal{F},\Lambda)$ for the matter fields, our initial step will be to use the Einstein equations and extract the Lagrangian by requiring that they yield as solution the metric described in Eq.~\eqref{eq:S2:metric} with the corresponding metric function of Eq.~\eqref{eq:f(r)}. In what follows the derived Lagrangian will be a function of the radial coordinate, but it may equivalently be expressed in terms of the field strength with Eq.~\eqref{eq:F(r)} providing us with the connection between these two expressions/representations. Regardless of its specific form, the corresponding energy-momentum tensor (EMT) can be found after varying the action with respect to $g_{\mu\nu}$, which leads to the Einstein equations, 
	\begin{align}
		G_{\mu\nu}=8\pi T_{\mu\nu}\label{eq:ein-eqs}
	\end{align}
    with an EMT of the form
	\begin{align}
		T_{\mu\nu}=-4\frac{\partial \mathcal{L}}{\partial \mathcal{F}}F^{\alpha}_{\nu}F_{\mu\alpha}+\mathcal{L}g_{\mu\nu}.\label{eq:EMT-NED}
	\end{align}   
	For our case, the identification of the Lagrangian requires using the following components of the Einstein tensor $G_{\mu\nu}$,
		\begin{align}
			G_{00}&=\frac{f(r)}{r^2}\left(1-f(r)-rf'(r)\right),\\
			G_{11}&=\frac{-1+f(r)+rf'(r)}{r^2f(r)},\\
			G_{22}&=\frac{1}{2}r\left(2f'(r)+rf''(r)\right),\label{eq:Gab}
		\end{align}
	along with the corresponding EMT components on the right-hand side of Eq.~\eqref{eq:ein-eqs}, which are calculated by using the EMT of Eq.~\eqref{eq:EMT-NED}, and are given by 
		\begin{align}
			T_{00}&=-f(r)\mathcal{L},\\
			T_{11}&=f(r)^{-1}\mathcal{L},\\
		    T_{22}&=r^2\mathcal{L}-4\frac{\partial\mathcal{L}}{\partial \mathcal{F}}\frac{Q_{m}^2}{r^2}.\label{eq:Tab}
		\end{align}
    In this derivation, we have assumed that the Lagrangian density shares the spherical symmetry of the geometry, and thus it is a function only of the radial coordinate $r$. Using the above equations, we find that the Lagrangian is given in terms of $r$ by 
	\begin{align}
		\mathcal{L}(r)=\frac{2\left(1-f(r)-rf'(r)\right)}{r^2}.\label{eq:L(r)-original}
	\end{align}
	Therefore, to generate the regular geometry we want to study in this paper, we need to substitute the metric function of Eq.~\eqref{eq:f(r)} in Eq.~\eqref{eq:L(r)-original},
    and arrive at the following Lagrangian:
    \begin{align}
    	\mathcal{L}(r)=\frac{12ml}{(r+l)^4}+2\Lambda,\label{eq:L(r)}
    \end{align}
    where it is evident that in the absence of the minimal length scale we are led to $\mathcal{L}(r)=2\Lambda$ as is required for consistency. The mass $m$ is the Komar mass \cite{K:63} of the metric,\footnote{When it comes to de Sitter spacetime, caution must be taken with this interpretation. This spacetime does not admit a globally timelike Killing vector field, so the construction of conserved charges is problematic \cite{BBM:02, B:02, ANS:11}. This is due to the presence of a cosmological horizon leading to a spacelike character of the Killing vector $\partial_{t}$, assocciated with time translations at distances larger than the cosmological horizon.} and we can write it as a function of the outer horizon radius $r_h$ and the minimal length scale $l$ by using the condition $f(r_h)=0$, which leads to 
    \begin{align}
    	m=\frac{(r_h+l)^3(3-\Lambda r^2_h)}{6r^2_h}.\label{eq:m}
    \end{align}
    The root $r_h$ corresponds to the largest root of the equation $f(r)=0$, except in the case of a de Sitter spacetime which admits one more horizon, the cosmological one, at larger distance in comparison with the outer horizon of the black hole.
    
    The source of this RBH is a particular case of the sources presented in Ref.~\cite{FW:16} whose Lagrangian density, as a function of the field strength $\mathcal{F}$, is given by\footnote{To avoid issues with the notation in this article we have replaced the parameter $a$ used in Ref.~\cite{FW:16} with the letter $\sigma$.} 
    \begin{align}
    	\mathcal{\tilde{L}}(\mathcal{F})=\frac{4\mu}{\sigma}\frac{\sigma \mathcal{F}}{\left(1+(\sigma\mathcal{F})^{1/4}\right)^{\mu+1}},\label{eq:Lt(F)}
    \end{align}
    which leads to a geometry of the form 
    \begin{align}
    f(r)=1-\frac{2\sigma^{-1}q^3r^{\mu-1}}{(r+q)^{\mu}},
    \end{align}
    in an asymptotically flat spacetime ($\Lambda=0$) with the magnetic charge given by 
    \begin{align}
    	Q_{m}=\frac{q^2}{\sqrt{2\sigma}}.
    \end{align}
    The spacetime geometry analyzed in this article can be generated by the following choice of parameters
    \begin{align}
    	\mu=3,\quad q=l, \quad \sigma=\frac{l^3}{m},\label{eq:sigma}
    \end{align}
    which lead to a magnetic charge 
    \begin{align}
    	Q_{m}=\sqrt{\frac{ml}{2}},\label{eq:Qm}
    \end{align}
     and the Lagrangian density given by Eq.~\eqref{eq:L(r)} once we use Eq.~\eqref{eq:F(r)}, which provides the link from one representation to another. Two brief comments are in order; firstly, the Lagrangian of Eq.~\eqref{eq:Lt(F)} has a Maxwell weak-field limit, which can be seen by performing an expansion around $\mathcal{F}$, and leads to 
     \begin{align}
     	\tilde{\mathcal{L}}(\mathcal{F})=4\mu \mathcal{F}+\cO{(\mathcal{F}^{
     	5/4})}.
     \end{align}
     This is a different feature of this subclass of NED Lagrangians in comparison with the ones used to generate other well-known models such as Bardeen or Hayward, which exhibit stronger weak-field limits than the Maxwell theory. To make this statement more precise, one finds using the sources described in Ref.~\cite{FW:16} that for the Bardeen model the weak-field limit is $\cO{(\mathcal{F}^{5/4})}$, whereas for the Hayward model is $\cO{(\mathcal{F}^{5/2})}$. 
     
      Secondly, it is evident from relation \eqref{eq:Qm} that the existence of the magnetic charge is tied to the presence of the minimal length scale. This means that in the case of vanishing charge, the minimal length vanishes as well and we retrieve the singular Schwarzschild geometry. Conversely, if we maintain a fixed nonzero magnetic charge, as we will in the study of phase transitions in the canonical ensemble, the minimal length persists as a fundamental feature. 
      
      Having established the matter Lagrangian that serves as a source for this theory, we can now proceed with the study of thermodynamics. We can approach this either using Hamiltonian methods \cite{ZG:18} or Euclidean path integral methods. Both approaches lead to congruent thermodynamic quantities, in the regimes where they can be directly compared, which is a prerequisite for consistency. A detailed analysis on the consistency of these approaches has been conducted in Ref.~\cite{SS:23} using as example the Hayward model. Without loss of generality, we restrict ourselves to the Euclidean path integral method described in the following section.

    \section{Thermodynamics with Euclidean action}\label{sec:thermodynamics}
    
    In spite of its inherent technical issues, the path integral has been firmly established as an invaluable tool for our understanding of gravity beyond the classical realm.  Aside from formally defining a quantum theory of gravity, it provides us with a powerful way to study gravitational thermodynamics, even in spacetimes that do not admit a straightforward definition of temperature, as for example the de Sitter case, which we will analyze later on. In this application, we exploit the relation between the classical Euclidean action $I_{E}$ and the quantum mechanical partition function $\mathcal{Z}$, 
    \begin{align}
    	F=-T\log{\mathcal{Z}}\approx TI_{E},
    \end{align} 
    where $F$ is the free energy of the system and $T$ the temperature. The above relation holds in the semiclassical approximation. We are concerned with the interaction of quantum matter fields with the presumably quantum gravitational field. Thus the path integral measure should include both the metric associated with the gravitational field along with the matter fields. At leading order in $m_{p}/M$, with $m_{p}$ representing the Planck mass, the matter fields do not contribute to the path integral. Consequently, the measure in use will solely originate from topologically distinct metrics. With all of these assumptions, the partition function can be approximated as 
    \begin{align}
    	\mathcal{Z}=\int_{g(0)}^{g(\tilde{\tau})}\mathcal{D}[g]e^{-I_{E}[g]/\hbar},
    \end{align}
    where $\tilde{\tau}$ is the periodicity. One further approximation can be made by considering only the leading contribution to the integral, which comes from metrics that are classical solutions to the equations of motion, namely those for which $\delta I_{E}[g_{cl}]=0$. This is the saddle point approximation, in which 
    \begin{align}
    	\int_{g(0)}^{g(\tilde{\tau})}\mathcal{D}[g]e^{-I_{E}[g]/\hbar}\approx e^{-I_{E}[g_{cl}]/\hbar}.
    \end{align}
    Therefore, we conclude that 
    \begin{align}
    	\mathcal{Z}\approx e^{-I_{E}[g_{cl}]/\hbar} \Rightarrow F\approx TI_{E}. \label{eq:Z:approx}
    \end{align}
     This can be regarded as the zero loop approximation to the full partition function, which only includes the dominant contribution from the gravitational field. Using the partition function, we can define thermodynamic quantities using well-known formulas from statistical mechanics. Using the approximation \eqref{eq:Z:approx}, one can determine the internal energy and entropy of the system through the Euclidean action with the following formulas
     \begin{align}
     	E=\frac{\partial I_{E}}{\partial \beta}, \quad S=\beta \frac{\partial I_{E}}{\partial \beta} -I_{E},\label{eq:ES-Ie}
     \end{align}
     where $\beta$ denotes the inverse temperature $T^{-1}$. This approach, in the case of black holes, was first developed by Gibbons and Hawking \cite{GH:77}, and analyzed further by York in Ref.~\cite{Y:86}.
    
     Having introduced the basic notation and methodology for the Euclidean path integral approach, we start our analysis with the calculation of the proper Euclidean action. For the model under consideration the total reduced action\footnote{We use the term ``reduced" action to distinguish it from the original general form, i.e., before any integration has been carried out.} $I_{r}$ is given by
    \begin{align}
    	I_{r}=I_{EH}+I_{GHY}+I_{M}+I_{EMB}-I_{0}.
    \end{align} 
    We will proceed with the explanation and calculation (see Appendix \ref{sec:app:Ir}) of each term separately and then combine them to arrive at the final expression for the total reduced action. We start with the Einstein-Hilbert action $I_{EH}$. We do not consider the cosmological constant in this action term since it is already implemented in the matter term $I_{M}$ of the action. We have that 
    \begin{align}
    	I_{EH}=-\frac{1}{16\pi}\int d^4x\sqrt{g} R,\label{eq:Ieh-initial}
    \end{align}
    where $R$ is the Ricci scalar for the Euclidean spacetime geometry. The next term of the action is the Gibbons-Hawking-York term \cite{GH:77, Y:86} that is introduced when we perform an integration over a region of spacetime bounded by a hypersurface $\partial M$. This term of the action is computed at $r=r_c$ and is given by 
    \begin{align}
    	I_{GHY}=\frac{1}{8\pi}\int_{\partial M}\sqrt{k}K, \label{eq:GHY-initial}
    \end{align}
    where $K$ is the trace of the extrinsic curvature.
   
     The matter part of the action which, as we saw in the derivation of the source of this geometry in Sec.~\ref{sec:S2 metric}, contains the cosmological fluid part as well and is given by
     \begin{align}
     I_{M}=\frac{1}{16\pi}\int d^4x \sqrt{g}\mathcal{L}(r),\label{eq:Im-initial}
     \end{align}
     where $\mathcal{L}$ is the Lagrangian density described by Eq.~\eqref{eq:L(r)}. 
     
     The next action term is the electromagnetic boundary term, which is introduced to keep the magnetic charge fixed since the thermodynamic analysis of this paper will be restricted to the canonical ensemble \cite{BBWY:90}. For an NED theory this term is given by
     \begin{align}
     	I_{EMB}=-\frac{1}{16\pi}\int_{\partial M}\sqrt{k}\left(\frac{\partial \mathcal{L}}{\partial \mathcal{F}}\right)F^{\mu\nu}n_{\nu}A_{\mu},
     \end{align}
     where $n_{\nu}$ is the unit normal vector to the boundary $\partial M$ \cite{MK:21,MK:22}. This term will vanish since we are integrating over a time slice of the spacetime and the only nonvanishing components of the electromagnetic tensor $F_{\mu\nu}$ are $F_{23}=-F_{32}$, as can be seen from Sec.~\ref{sec:S2 metric}.
     
      The final term of the action $I_{0}$ corresponds to a subtraction term introduced to regularize the action. It is chosen such that the total action vanishes when a black hole is not present, i.e.,~$m=0$. Before we explicitly write the final expression for the reduced action, we point out that in order to properly compare a black hole spacetime to a spacetime without a black hole, we need to match the boundaries (cavity) in these two spacetimes leading to a new parameter $\beta$ in the background action instead of $\beta_{h}$ \cite{BBWY:90, CV:03}.
     Once we combine all of the nonvanishing parts of the action, namely Eqs.~\eqref{eq:Ieh}, \eqref{eq:Ighy}, \eqref{eq:Im}, and \eqref{eq:I0}, we are led to the final expression for the reduced action,
     \begin{widetext}
     \begin{align}
     	I_{r}=-\pi r^2_h+\frac{\beta r_c}{3}\left(3-r^2_c\Lambda-\sqrt{3-r^2_c\Lambda}\sqrt{3-r^2_c\Lambda +\frac{r^2_c(r_h+l)^3(-3+r^2_h\Lambda)}{(r_c+l)^3r^2_h}}\right).\label{eq:Ir}
     \end{align}
     \end{widetext}   
     Following the methodology in Ref.~\cite{BBWY:90}, we now proceed with the computation of the proper thermodynamic quantities. For the calculation of the temperature, we need to identify the stationary points of the action. This involves extremizing it with respect to the horizon radius $r_h$ and leads to the equation, 
     \begin{align}
     	\frac{\partial I_{r}}{\partial r_h }=0,
     \end{align}
     which can be used to determine $\beta$. The solution $\beta$ is a function of the cavity radius $r_c$, the horizon radius $r_h$, the minimal length scale $l$, and the cosmological constant $\Lambda$. Therefore, the temperature for this black hole will be $T=\beta^{-1}(r_h,r_c,l,\Lambda)$ and it is given by
     \begin{align}
     	T=\frac{r^3_c(r_h+l)^2\sqrt{3-r^2_c\Lambda}(2l-r_h+r^3_h\Lambda)}{4\pi (r_c+l)^3 r^4_{h} \mathcal{X}},\label{eq:Tds}
     \end{align}
     where $\mathcal{X}$ is given by the following, rather long expression
     \begin{widetext}
     \begin{align}
     	\mathcal{X}=\sqrt{\frac{-(r_c-r_h)(9l^2r_cr_h-3r^2_cr^2_h+3l^3(r_c+r_h)+r^2_cr^2_h(3l^2+r^2_c+r_c r_h+r^2_h+3l(r_c+r_h))\Lambda)}{(r_c+l)^3r^2_h}}.
     \end{align}
     \end{widetext}
     We proceed with the calculation of the remaining thermodynamic quantities in the same manner as in Ref.~\cite{BBWY:90}. For the minimal length scale $l$, the conjugate potential that appears in the first law of black hole thermodynamics and the Smarr formula is given by 
     \begin{align}
     	\Phi=\frac{1}{\beta}\frac{\partial I_{r}}{\partial l},
     \end{align} 
     and an explicit calculation yields 
     \begin{align}
     	\Phi=\frac{r^3_c(r_c-r_h)(l+r_h)^2\sqrt{3-r^2_c\Lambda}(3-r^2_h\Lambda)}{2(l+r_c)^4r^2_h\mathcal{X}}.\label{eq:phi}
     \end{align}    
     In the presence of a cosmological constant we have a relevant pressure/tension (depending on the sign) arising from it, which is defined via $P=-\Lambda/8\pi$. In a similar way, we find the conjugate variable to the pressure/tension $P$, which will correspond to the thermodynamic volume $V$. It is defined by 
     \begin{align}
     	V=\frac{-8\pi}{\beta}\frac{\partial I_{r}}{\partial \Lambda}, \label{eq:Vth}
     \end{align}
     which leads to a rather long expression that does not provide any insight for the reader, so it is given in Appendix \ref{sec:app:therm-potentials}. We emphasize here that we are working in the extended phase space, allowing for variations of $\Lambda$ in the first law \cite{KMT:17}. Motivation for permitting the variation of the cosmological constant and the minimal length scale is given in Sec.~\ref{sec:1st-law}.
     
     There is one more relevant potential, denoted as $\lambda$, that needs to be computed.  It is conjugate to the area of the cavity $A_c=4\pi r^2_c$. We will interpret this term as the surface pressure or tension, depending on the sign, resulting from the presence of the cavity. We have that 
     \begin{align}
     	\lambda=\frac{1}{\beta}\frac{\partial I_{r}}{\partial A_c}.
     \end{align} 
     In the above calculation, we first replace $r_c\rightarrow \sqrt{A_c/4\pi}$ and then take the derivative. After that, we substitute $A_{c}$ with $4\pi r^2_c$ and we have the conjugate potential $\lambda$ as a function of the cavity radius. The full expression is provided in Appendix \ref{sec:app:therm-potentials}. In the absence of a cavity this term will not contribute to the first law or the Smarr formula, as we will see in the AdS and Minkowski cases later on. 
       
     Now that we have completed the computation of the conjugate potentials, we can proceed with the calculation of two highly significant quantities for thermodynamics, namely the mean thermal energy $E$ and the entropy $S$. These calculations are carried out using Eq.~\eqref{eq:ES-Ie}. The mean thermal energy plays the role of the internal energy in the first law and is given by 
     \begin{align}
     	E=\frac{\partial I_{r}}{\partial \beta},\label{eq:E-dIr}
     \end{align} 
     which in the presence of an isothermal cavity yields 
     \begin{widetext}
     \begin{align}
     	E=\frac{r_c}{3}\left(3-r^2_c\Lambda-\sqrt{3-r^2_c\Lambda}\sqrt{3-r^2_c\Lambda+\frac{r^2_c(l+r_h)^3(-3+r^2_h\Lambda)}{(l+r_c)^3r^2_h}}\right).\label{eq:energy}
     \end{align} 
     \end{widetext}
     The entropy is calculated from the Euclidean action from the relation
     \begin{align}
     	S=\beta \frac{\partial I_{r}}{\partial \beta}-I_{r},\label{eq:S-dIr}
     \end{align}
     which leads to the well-known result due to Bekenstein and Hawking,
     \begin{align}
     	S=\frac{A}{4}=\pi r^2_h, \label{eq:entropy}
     \end{align}
     with $A$ being the horizon area. The above quantities define a proper Smarr relation with the appropriate scaling arguments \cite{KRT:09}, which is written as 
     \begin{align}
     	E=2TS+\Phi l+2\lambda A_{c}-2PV.
     \end{align}
     The scaling arguments correspond to the following substitutions:
     \begin{align*}
     	E\rightarrow cE,\quad l\rightarrow cl,\quad r_h\rightarrow cr_h,\quad \Lambda\rightarrow c^{-2}\Lambda,\quad  S\rightarrow c^2S,
     \end{align*}	
     \begin{align}
     	T\rightarrow c^{-1}T,\quad \Phi\rightarrow \Phi,\quad\lambda\rightarrow c^{-1}\lambda,\quad V\rightarrow c^{3}V. \label{eq:scal-arg}
     \end{align}
     The first law of black hole mechanics is given by 
     \begin{align}
     	dE=TdS+\Phi dl+\lambda dA_c+VdP.
     \end{align}
     We note that in the above expression of the first law $E$ is identified as the Komar mass in the absence of a cosmological constant, whereas if $\Lambda<0$ is present $E$ has to be interpreted as the enthalpy of the system \cite{KMT:17}. Intuitively, this means that now $E$ will correspond to the internal energy $M$ of the system (Komar mass) in addition to the energy $PV$ needed to displace the vacuum energy from its environment, or in other words, to embed the black hole in the spacetime with the negative cosmological constant. We note here that it is trickier to give, in an analogous way, thermodynamic/physical interpretations of quantities in de Sitter spacetime due to the presence of the cosmological horizon.   
     
     We will use the derived thermodynamic quantities with their appropriate limits, when possible, in the following section where we will analyze the phase structure of the model described by Eq.~\eqref{eq:f(r)}.
     
     \section{First law of black hole thermodynamics, Smarr formula and phase structure}\label{sec:1st-law}
	The first law of black hole mechanics derived in Ref.~\cite{BCH:73} has been proven to hold in any theory of gravity arising from a diffeomorphism-invariant Lagrangian \cite{W:93,IW:94}. Extension of the first law for the case of NED was first studied in Ref.~\cite{R:03}, although the Smarr formula was not satisfied. This problematic behavior was fixed later on in Ref.~\cite{ZG:18}, where the Bardeen black hole and the Born-Infeld theory \cite{BI:34} were studied, by appropriate accounting of the extra parameters of the theory. The mathematical formulation of the first law in an asymptotically flat spacetime in the case of NED theory is 
	\begin{align}
		dM=\mathcal{T}dS+\Phi_{e} dQ_{e}+\Psi_{H}dQ_{m}+\sum_{i}K_{i}d\beta_{i},\label{eq:1st law-gen}
	\end{align}
	where $M$ is the Komar mass, $\mathcal{T}$ is the temperature given by the surface gravity $\kappa$ as $\mathcal{T}=\kappa/2\pi$, $S$ is the entropy, $\Phi_{e}$ and $\Psi_{H}$ are the electric and magnetic potentials associated with the electric and magnetic charge $Q_{e}$ and $Q_{m}$, respectively, and the terms $K_{i}$ correspond to potentials linked to the extra parameters $\beta_{i}$ of the theory. 
	 Using Hamiltonian methods it is possible to calculate these potentials and with appropriate change of variables we arrive at the thermodynamic quantities arising naturally from the Euclidean action \cite{SS:23}. 
	
    In the following, the minimal length will be considered as a fundamental parameter and we will allow its variation in a region that will admit a regular black hole geometry (although not extremal). Before we proceed to the analysis of the first law, Smarr formula, and phase structure for each case of spacetime, it is useful to provide some motivation for allowing parameters such as the minimal length $l$ and the cosmological constant $\Lambda$ to vary and therefore appear in the first law. In general, we expect that there are more fundamental theories in which coupling constants or the cosmological constant are not predetermined but on the contrary emerge from vacuum expectation values allowing for possible variations thereof. Consequently, it is physically justified to include their variation in the first law \cite{GKK:96,CM:95}. In the presence of a cosmological constant or extra parameters of the theory, i.e.~minimal length in the majority of RBHs, one can see that the Smarr relation is not satisfied, and therefore we have to include a variation of the extra parameters in the first law to ensure consistency \cite{KRT:09,B:05}.
	
	We provide some additional motivation for the varying minimal length. Usually this length is associated with a fundamental scale set by a quantum gravity theory, i.e.~the Planck scale \cite{BCSV:19,G:95}. However, studies that take into account backreaction in the black hole interior lead to an evolving minimal length $l$ \cite{BBCRG:21,BBCRG:22,CRFLV:23}, therefore permitting its variation in the first law. 
	
	We now proceed with the study of the thermodynamics of the RBH in spacetimes with and without $\Lambda$. In the case of a negative cosmological constant, i.e.,~AdS spacetime, the thermal equilibrium is established naturally. The boundary of the AdS is timelike and this allows massless particles to reach it in finite proper time. The commonly accepted approach involves enforcing reflecting boundary conditions, effectively treating AdS spacetime as closed. As a result, massless particles will be reflected towards the center upon reaching $r=\infty$. The same behavior applies to massive particles, but this is because of the attractive nature of the potential when $\Lambda<0$. This feature of the spacetime allows for sufficiently large black holes to reach a state where the rate of outgoing radiation matches the one of the reflected (ingoing) radiation, rendering the black hole stable. This leads to an equilibrium state with one temperature $T$ which is the temperature of the black hole, or equivalently, of the radiation. 
	
	To enforce thermal equilibrium in spacetimes such as de Sitter and Minkowski, we need to introduce an isothermal cavity \cite{KS:16,SM:18,SM:19,HHMS:20,SFM:21}, otherwise this is not possible. In the de Sitter case, the presence of the cosmological horizon which, in general, radiates at a different temperature than the black hole horizon, leads to an observer situated inbetween them to observe a nonequilibrium state because of this temperature difference \cite{S-essay:23}.
	
	While studying phase transitions, we will consider the order parameter of the theory to be the horizon radius $r_h$, which represents the size of the black hole. It is worth pointing out that in this case the value $m=0$ represent the empty spacetime, which is equivalent to $r_h=0$ and $l=0$ as we saw in Sec.~\ref{sec:S2 metric}. We proceed with a separate study for each spacetime case now.
	
	\subsection{Embedding in anti-de Sitter}\label{sec:PT-AdS}
    
    The AdS spacetime does not admit any other horizon apart from the black hole's inner and outer ones and therefore we are permitted to study the thermodynamic properties in the absence of a cavity, i.e.,~taking the cavity radius $r_c$ to infinity. Based on this, we proceed with the computation of the appropriate thermodynamic quantities. We start with the temperature which is calculated, by taking the limit $r_c\rightarrow \infty$ of Eq.~\eqref{eq:Tds} to be  
    \begin{align}
    	T_{AdS}=\lim_{r_c\rightarrow \infty}T=\frac{(l+r_h)^2(r_h-2l-r^3_h\Lambda)}{4\pi r^4_h}.\label{eq:Tads}
    \end{align}
	We should emphasize that this is an effective temperature that arises directly from the Euclidean path integral approach and differs from the ordinary/conventional definition of the temperature linked to the surface gravity $\kappa$\footnote{We point out that in the form of the first law described by Eq.~\eqref{eq:1st law-gen} the temperature $\mathcal{T}$ is given by the surface gravity. In our case, except for the magnetic charge $Q_m$, we have an additional parameter $\beta_{i}=\sigma$, which appears in the NED Lagrangian of Eq.~\eqref{eq:Lt(F)}. If we treat as fundamental thermodynamic variables the magnetic charge $Q_{m}$ and the parameter $\sigma$, instead of the minimal length scale $l$, then the first law, in the AdS case, takes the form $dM=\mathcal{T}dS+\Psi_{H}dQ_{m}+K_{\sigma}d\sigma+VdP$. Under this assumption, using the Euclidean path integral formalism, the extremization of the action with respect to the horizon radius leads to the temperature being the surface gravity and the entropy remains the Bekenstein-Hawking entropy. For a detailed derivation of this result and a comparison between considering $Q_m$ and $\sigma$ as fundamental variables versus the minimal length scale $l$ we refer the reader to Appendix \ref{sec:app:action-Qm}.}. To illustrate this point clearly, we provide the expression for this temperature as
	\begin{align}
		\mathcal{T}=\frac{\kappa}{2\pi}=\frac{r_h-2l-r^3_h\Lambda}{4\pi r_h(r_h+l)}.\label{eq:Tsg}
	\end{align} 
	An elegant relationship between these two temperatures becomes apparent, demonstrating that the effective temperature is slightly higher due to the regularization of the singularity through the introduction of the minimal length. This relation is 
	\begin{align}
		T_{AdS}=\left(1+\frac{l}{r_h}\right)^3\mathcal{T}, \label{eq:temp-comparison}
	\end{align}
	which leads to $T_{AdS}>\mathcal{T}$, although by a small amount since we expect in general $l\ll r_h$.\footnote{We emphasize that this higher effective temperature is not unique to our scenario. It also manifests in singular geometries, as the one explored in Ref.~\cite{LHK:23}. This ``modified" temperature should be attributed to the presence of matter. However, in the case we are considering, the matter fields are tied to the singularity regularization allowing us to make a direct connection between this increased temperature and the absence of singularity.}
	
	 \begin{figure*}[!htbp]
		\centering
		\begin{tabular}{@{\hspace*{-0.0\linewidth}}p{0.45\linewidth}@{\hspace*{0.05\linewidth}}p{0.45\linewidth}@{}}
			\centering
			\subfigimg[scale=0.85]{(a)}{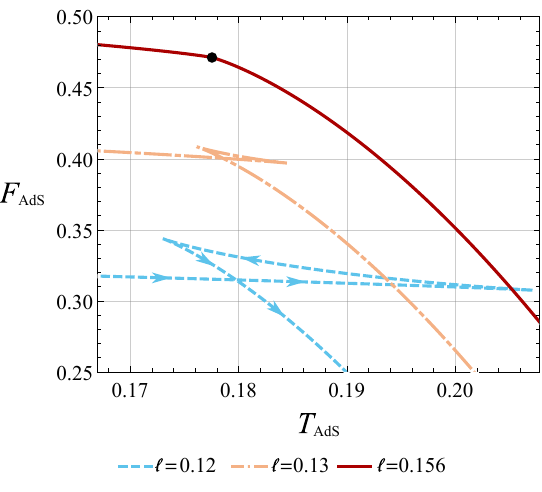} &
			\subfigimg[scale=0.85]{(b)}{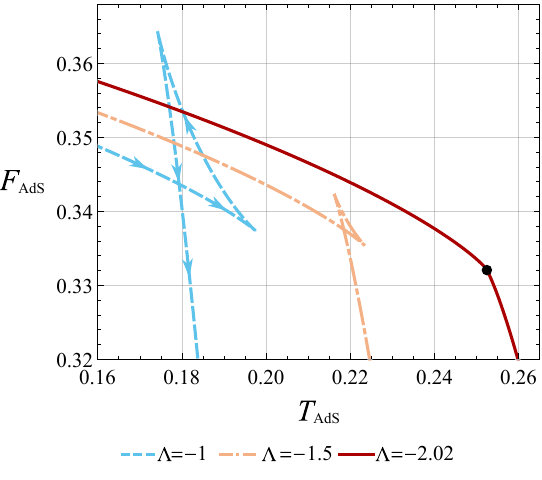}
		\end{tabular}
		\caption{(a) Free energy $F_{AdS}$ as a function of the temperature $T_{AdS}$ for a constant value of the cosmological constant $\Lambda=-1$ with varying minimal length $l$; (b) Free energy $F_{AdS}$ as a function of the temperature $T_{AdS}$ for a constant value of the minimal length $l=0.11$ with varying cosmological constant $\Lambda$; In both cases, the arrows denote the direction in which the horizon radius $r_h$ increases. For the sake of readability, we omit the arrows in the remaining curves, although the same pattern is implied.} \label{fig:PT-AdS}
	\end{figure*}
	
	 We will now comment on the possible reasons behind this distinction between surface gravity and effective temperature. One possible explanation lies in the definition of the temperature as the surface gravity which is derived solely from the kinematic aspects of the spacetime geometry. Gravitons also contribute to the Hawking radiation and therefore their kinematic properties will play a significant role in the computation of the temperature/surface gravity. It is usually assumed that they move at the speed of light, although, it has been shown in Ref.~\cite{HLJV:21} that when we have scalar-tensor theories of gravity the propagation speed of gravitons is altered and they can move on timelike/spacelike trajectories. This could lead to a justified modification of the temperature. This reasoning was proposed as a possible explanation in Ref.~\cite{LHK:23} for the case of 4D scalar-tensor Einstein-Gauss-Bonnet gravity. An analogous phenomenon might be present in RBHs, as some of them can also be generated in the same framework. Recently, the Hayward metric has been derived in this context in Ref.~\cite{NN:23}. Another equally possible explanation relies in the interpretation of quantities introduced in the first law of black hole thermodynamics. When we consider variations of the energy in the first law, this energy represents the total energy of the spacetime, including the matter which may or may not be thermalized with respect to the Killing time. In this variation, maintaining the entropy of the total matter content as the geometric entropy necessitates a modification of the temperature, which is the one derived using the Euclidean path integral approach. However, when matter content is absent, for example by taking $l\rightarrow 0$ in Eq.~\eqref{eq:temp-comparison}, the two temperature definitions coincide. 
			
	 We continue with the calculation of the conjugate potential for $l$, which is derived by taking the limit $r_c\rightarrow \infty$ of Eq.~\eqref{eq:phi}, and find 
	\begin{align}
		\Phi_{AdS}=\frac{(l+r_h)^2(3-r^2_h\Lambda)}{2r^2_h}.\label{eq:phi-ads}
	\end{align}
	Similarly, we derive the thermodynamic volume $V_{AdS}$ by taking the limit $r_{c}\rightarrow \infty$ of Eq.~\eqref{eq:Vth}, which leads to
	\begin{align}
		V_{AdS}=\frac{4\pi}{3}(r_h+l)^3. \label{eq:therm-vol-ads}
	\end{align}
	We observe here the well-known result of the thermodynamic volume being different than the geometric one, while they happen to coincide when the minimal length scale vanishes, i.e.,~$l=0$, which corresponds to the Schwarzschild-AdS case. The mean thermal energy is given by relation \eqref{eq:energy} taking the limit $r_{c}\rightarrow \infty$, and it is 
	\begin{align}
		E_{AdS}=\frac{(l+r_h)^3(3-r^2_h\Lambda)}{6r^2_h}.\label{eq:energy-ads}
	\end{align}
	We see that the thermal energy is the same as the Komar mass that someone will calculate by solving the equation $f(r_h)=0$ given by Eq.~\eqref{eq:m}. This is not the case in de Sitter spacetime as we can see from Eq.~\eqref{eq:energy}. Lastly, as the cavity is positioned at infinity, it does not contribute any pressure, so there is no term $\lambda$ present in the first law or the Smarr formula.

	Therefore, in the AdS spacetime for this RBH, we have that the first law of black hole thermodynamics is given by 
	\begin{align}
		dE=T_{AdS}dS+\Phi_{AdS}dl+V_{AdS}dP.
	\end{align}
	We can also see that the Smarr formula is satisfied and it can be written in the form 
	\begin{align}
		E_{AdS}=2T_{AdS}S+\Phi_{AdS}l-2PV_{AdS},
	\end{align}
    where the scaling arguments \eqref{eq:scal-arg} hold. We note that the above Smarr relation is characterized by linearity even though the vector potential does not share the symmetry of the spacetime\footnote{The fourth component of the vector potential $A^{\mu}$ depends on the angle $\theta$ as can be seen from Eq.~\eqref{eq:Am}.}. The symmetry inheritance of the spacetime from the fields is a sufficient condition for linearity \cite{GS:17}, but in this case we see that this condition is not satisfied although linearity is preserved. 
	
	Since we have calculated the proper thermodynamic quantities for AdS, we can now proceed with the study of the phase structure of this geometry. We analyze it in the canonical ensemble, but before that, we need to calculate the crucial quantity for this study, which is the Gibbs free energy defined as 
	\begin{align}
		F_{AdS}=E_{AdS}-T_{AdS}S.
	\end{align}
	After using Eqs.~\eqref{eq:energy-ads}, \eqref{eq:Tads} and \eqref{eq:entropy}, we have that 
	\begin{align}
		F_{AdS}=\frac{(l+r_h)^2(12l+3r_h+r^2_h\Lambda(r_h-2l))}{12r^2_h}.
	\end{align}
	The equilibrium state of the thermodynamic system corresponds to the global minimization of the free energy. To properly study the phase structure, we need to study the free energy as a function of temperature by drawing parametric plots using the outer horizon radius as a parameter. Figure~\ref{fig:PT-AdS} reveals two distinct yet quite similar behaviors. The distinction comes from the fact that we vary different parameters in Figs.~\ref{fig:PT-AdS}(a) and \ref{fig:PT-AdS}(b), i.e.,~the minimal length and the cosmological constant, respectively. We observe from Fig.~\ref{fig:PT-AdS} a characteristic swallowtail behavior. For a fixed value of the cosmological constant $\Lambda=-1$, there is a critical value for the minimal length $l_{c}\approx 0.156$ below which there is a first-order phase transition from a small to a large black hole, while above this critical value, no phase transition occurs. The same behavior is exhibited when we vary the cosmological constant, where the critical value is $\Lambda_{c}\approx -2.02$ below which no phase transition occurs, and above there is a small to large black hole first-order phase transition. The parameter that changes here is the horizon radius $r_h$: it increases from left to right along the near horizontal lines, which is the same direction as the increasing temperature. The direction of increasing radius is also indicated with arrows for the dashed light blue curve. Starting from low temperatures on this curve and increasing it gradually, we reach the crossing point. At this point, the system chooses the direction that minimizes the free energy which is downwards, indicating that from a small horizon radius, we go straight to a large horizon radius, i.e.,~a small to large phase transition. At the critical point indicated by a black dot in Fig.~\ref{fig:PT-AdS}, the transition becomes second order. 
	
	\begin{figure}[!htbp]
		\hspace*{-1cm}
		\includegraphics[scale=0.85]{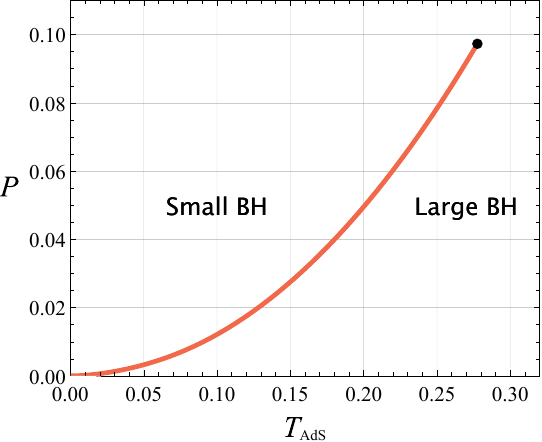}
		\caption{Coexistence line for the RBH model in AdS for a fixed minimal length scale $l=0.1$. The orange line represents points where first-order phase transitions occur, and the black dot represents the termination of the swallowtail with a second-order phase transition occurring at the critical point $\left(P^{(c)}, T^{(c)}_{AdS}\right)=\left(0.2776,0.0973\right)$. The origin $(0,0)$ of the diagram does not belong to the curve since the pressure must be positive in AdS. } 
		\label{fig:coex-line}
	\end{figure}
	
	Therefore, in the extended phase space, this RBH model embedded in AdS exhibits the same critical behavior as ordinary fluid systems \cite{KMT:17}. Combining all the crossing points, a swallowtail is created and it closes at the critical point. These crossing points represent states where the small and large phases coexist. Combining all of these points, we can illustrate the coexistence line in a $P-T$ diagram which terminates at the critical point, as can be seen from Fig.~\ref{fig:coex-line}. At temperatures above $T^{(c)}_{AdS}$ and/or pressures above $P^{(c)}$ there is no difference between the two phases, so as the temperature increases we have a smooth transition from a small to a large black hole. This behavior is reminiscent of the liquid-gas transition.

	  The phase structure of this RBH is similar to the one encountered when studying the Reissner-Nordstörm metric in the canonical ensemble. There, the presence of a nonzero electric charge makes the line $F=0$, i.e.,~pure radiation, inaccessible, thus preventing the Hawking-Page transition from occurring. This happens because in the presence of a nonzero charge the black hole cannot evaporate completely. This situation is similar to our case, where in the canonical ensemble the magnetic charge is fixed, leading to the same behavior. The main difference here is that the magnetic charge is inherently linked to the presence of the minimal length, as seen from Eq.~\eqref{eq:Qm}, and therefore the singularity regularization. This suggests that there might be a strong connection between the presence of the singularity and the existence of a Hawking-Page transition. Similar behavior has been encountered before in the study of the Hayward black hole in Ref.~\cite{SS:23}. 
  
	  We proceed now with the calculation of the critical point and examine if we have the usual mean-field theory behavior encountered in other black hole cases in AdS \cite{KMT:17}. To do that, we write the pressure as a function of the temperature $T_{AdS}$ and the thermodynamic volume $V_{AdS}$. The volume expression, given in Eq.~\eqref{eq:therm-vol-ads}, is a monotonic function of the horizon size $r_h$, and we can freely use this parameter to calculate the critical points in its stead. Therefore, we write the pressure as a function $P\left(r_h, T_{AdS}\right)$ which is computed by solving Eq.~\eqref{eq:Tads} for $\Lambda$. The pressure is identified as $-\Lambda/8\pi$ and thus we conclude that 
	  \begin{align}
	  	P\left(r_h,T_{AdS}\right)=\frac{T_{AdS}}{2r_h\left(1+\frac{l}{r_h}\right)^2}-\frac{r_h-2l}{8\pi r^3_h}.\label{eq:pressure}
	  \end{align}
	  From the above relation, we can extract information about the reduced volume, i.e.,~the volume over the number of particles, which is the inverse coefficient of the linear term of the temperature \cite{landau}. Therefore, we conclude that the reduced volume is given by
	  \begin{align}
	  	v=2r_h\left(1+\frac{l}{r_h}\right)^2.\label{eq:reduced-volume}
	  \end{align}
	  The definition of the reduced volume arises naturally from the Van der Waals-like form of Eq.~\eqref{eq:pressure}, and here it is different than the usually assumed value $2r_{h}$ \cite{KMT:17}, although in the limit of vanishing minimal length we retrieve the usually defined quantity.
	  
	  \begin{figure}[!htbp]
	  	\hspace*{-0.3cm}
	  	\includegraphics[scale=0.85]{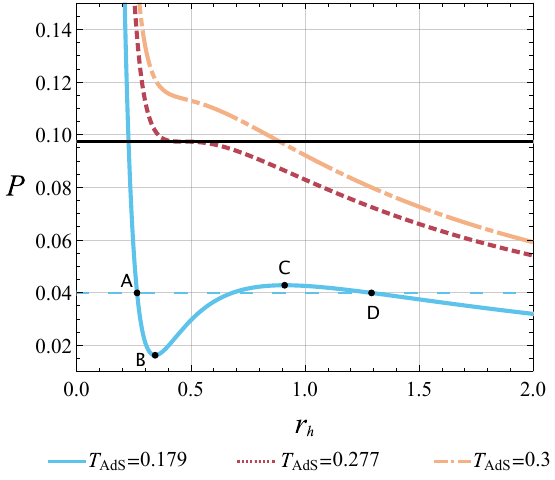}
	  	\caption{Isotherms of the equation of state $P\left(r_h,T_{AdS}\right)$ for the RBH model in AdS for a fixed value of the minimal length $l=0.1$ for various $T_{AdS}$. The solid horizontal black line represents the critical pressure $P^{(c)}\approx 0.973$, while the dashed horizontal light blue line depicts the coexistence line for the isotherm with temperature $T_{AdS}=0.179$.} 
	  	\label{fig:PV}
	  \end{figure}
	  
	  To calculate the critical point, we  need to find solutions for the following system of equations:
	  \begin{align}
	  	\frac{\partial P}{\partial r_h}=0, \quad \frac{\partial^2 P}{\partial r^2_h }=0.
	  \end{align}
	  Remarkably, analytic solutions for these equations are possible but because of their lengthy and complicated nature we omit them here and give instead approximate expressions. We find that 
	  \begin{align}
	  	r^{(c)}_h\approx 4.66433 l, \quad T^{(c)}_{AdS}\approx \frac{0.027758}{l}.
	  \end{align} 
	  Substituting these expressions into Eqs.~\eqref{eq:pressure} and \eqref{eq:reduced-volume} we find the respective critical values,
	  \begin{align}
	  	P^{(c)}\approx \frac{0.0009728}{l^2}, \quad v^{(c)} \approx 13.7574 l. \label{eq:crit-values}
	  \end{align}
	  For temperatures and pressures larger than the derived above critical values, phase transitions are not possible. This is also reflected later on in the study of the heat capacity.
	  
	  When studying thermodynamic systems, stability requirements dictate the presence of a positive heat capacity, and pressure to be monotonically decreasing function of the volume. We can see from Fig.~\ref{fig:PV} (orange isotherm) that above the critical temperature the pressure is a monotonically decreasing function of the horizon radius $r_h$, and therefore the thermodynamic volume $V_{AdS}$, due to their monotonically increasing relation. At the critical temperature (dark red line) we have a certain horizon radius for which $\partial P/\partial r_h=0$ and $\partial^2 P/\partial r^2_h=0$. Below the critical temperature, where phase transitions are possible, we have different monotonicity of the pressure with respect to the volume for different segments of the isotherm. The segments AB and CD correspond to the pressure being a decreasing function of the volume, and they represent metastable states (AB and CD correspond to superheated liquid and supercooled vapor in the liquid-gas case, respectively). The thermodynamic inequality $\partial P/\partial r_h<0$ is satisfied in these segments, but this is not the case in the segment BC where pressure is an increasing function of the volume, which signifies the violation of the thermodynamic inequalities and consequently thermodynamic instability. 
	  
	  The dashed horizontal light blue line in Fig.~\ref{fig:PV} corresponds to the coexistence pressure $P_{coex}$ derived by Maxwell's construction. In a $P-V$ diagram, the area enclosed between the isotherm and the line representing the coexistence pressure vanishes. This is a condition derived directly from the assumption of thermodynamic equilibrium between two phases which can be written as 
	  \begin{align}
	  	\int_{V_{A}}^{V_{D}}\left[P\left(r_h,T_{AdS}\right)-P_{coex}\right]dV=0.\label{eq:maxwell}
	  \end{align}
	  Here $V_{A}$ corresponds to the small black hole phase and $V_{D}$ to the large one. We can see by direct calculation that Eq.~\eqref{eq:maxwell} holds and that the coexistence pressure is the same as the one extracted from the $F-T$ diagram. We point out that proper calculation requires using the volumes rather than the radii of the small and large black holes. If we attempt to use the radii to perform the integration of the pressure, we are led to an inconsistency, namely that the coexistence pressure defined from Maxwell's construction does not coincide with the one found from the $F-T$ diagram.

	  We proceed now with the calculation of the critical ratio, which can be calculated using the critical values of the pressure, temperature, and reduced volume and find that
	  \begin{align}
	  	\frac{P^{(c)}v^{(c)}}{T^{(c)}_{AdS}}\approx 0.482197 \neq \frac{3}{8}.
	  \end{align}	  
	  This result indicates a deviation from the ``universal" mean-field theory critical ratio. On the other hand, if we consider the ordinary definition of the reduced volume $\tilde{v}=2r_{h}$ used in the majority of studies of black hole phase transition, we are led to a different critical ratio 
	  \begin{align}
	  	\frac{P^{(c)}\tilde{v}^{(c)}}{T^{(c)}_{AdS}}\approx 0.326968\neq \frac{3}{8},
	  \end{align}
	  which also signifies the departure from the usual value for the Van der Waals fluids, although there is a significant difference between the critical values calculated for different definitions of the volume. This is in contrast with what we have seen in Ref.~\cite{SS:23} and we attribute this deviation to the stronger deformation due to the minimal length in comparison with the Hayward model (See Table \ref{table:1}). 
	  
	   \begin{table}[!htbp]
	  	\def\arraystretch{2.5}
	  	\centering
	  	\begin{tabular}{|c|c|c|}
	  		\hline
	  		RBH model & Asymptotic behavior  & MFT ratio deviation \\
	  		\hline\hline
	  		Hayward & $\displaystyle 1-\frac{2m}{r}+\frac{4m^2l^2}{r^4}+\cO{(r^{-7})}$  & $0.016$\\
	  		\hline
	  		Bardeen & $\displaystyle 1-\frac{2m}{r}+\frac{3ml^2}{r^3}+\cO{(r^{-5})}$  &  $0.025$\\
	  		\hline
	  		Cadoni \textit{et al.} & $\displaystyle 1-\frac{2m}{r}+\frac{6ml}{r^2}+\cO{(r^{-3})}$     &  $0.107$\\
	  		\hline
	  	\end{tabular}
	  	\caption{In this table, the left column represents the RBH models we are considering, where the last model is the one analyzed in this paper. The middle column describes the asymptotic behavior where we have omitted the cosmological constant term. The right column describes the magnitude of the difference $|P^{(c)}v^{(v)}/T^{(c)}_{AdS}-3/8|$ between the critical ratios for these models with the usual mean-field theory (MFT) value $3/8$. In the calculation of the MFT ratio deviation, we have used only the reduced volume that appears as an inverse coefficient of the temperature in the Van der Waals-like equation of state, which should be considered the appropriate choice.}
	  	\label{table:1}
	  \end{table}
	  
	  It is interesting to compare the deviation from the value of the critical ratio of mean-field theory for a variety of RBH models that exhibit different corrections to the Schwarzschild asymptotic behavior. For this purpose, we consider the Hayward and Bardeen models\footnote{The thermodynamics of the Bardeen black hole is analyzed in the Appendix \ref{sec:app:Bardeen}.} along with the one analyzed in this paper.  We can see from Table \ref{table:1} that the deviation from the mean-field theory ratio becomes larger as the deformations of the asymptotic behavior become larger. This indicates that there is a connection between the way the singularity is smoothed out and the thermodynamic properties of black holes in AdS. It appears that the stronger the deformation from the Schwarzschild behavior using a minimal length scale, the larger the deviation from the mean-field theory critical ratio.

	   \begin{figure}[htbp]
	   	\hspace*{-1cm}
	  	\includegraphics[scale=0.8]{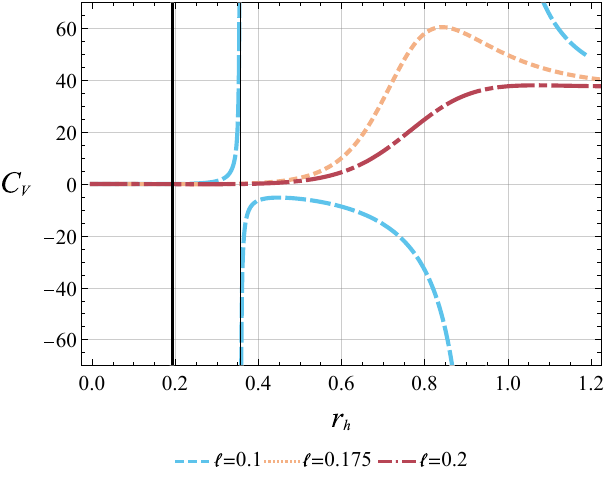}
	  	\caption{Heat capacity $C_{V}$ as a function of the horizon radius $r_h$ for constant value $\Lambda=-1$ of the cosmological constant. Different colors/types represent different values of the minimal length parameter $l$. The thin vertical black lines correspond to infinities/discontinuities of the heat capacity while the thick vertical black line represents the minimal radius enforced by the extremal limit for the case $l=0.1$.  } 
	  	\label{fig:Cv}
	  \end{figure}
	  
	  We now turn to the study of the thermodynamic stability of this RBH, where the critical quantity for this analysis will be the heat capacity at constant volume defined as
	  \begin{align}
	  	C_{V}=T\left(\frac{\partial S}{\partial T}\right)\bigg|_{V},
	  \end{align}
	  and it is calculated to be 
	  \begin{align}
	  	C_{V}=\frac{2\pi r^2_h(r_h+l)(2l-r_h+r^3_h\Lambda)}{\left(r^2_h-8 l^2+r^4_h\Lambda-lr_h(1+r^2_h\Lambda\right))}.\label{eq:Cv-ads}
	  \end{align}
	  We can see from Fig.~\ref{fig:Cv} that there are regions where the heat capacity is positive. They correspond to the thermodynamic stability of the RBH, whereas regions with the negative sign of $C_{V}$ represent thermodynamically unstable regions. Caution must be taken in our analysis since there is one additional constraint when we are studying RBHs. The smaller possible radius will be the one of the extremal black hole. In this case, the RBH will have one horizon \cite{MS:23} which is characterized by zero surface gravity since it is degenerate leading to zero temperature.  To identify this condition, we solve the following equation:
	  \begin{align}
	  	\kappa=\frac{f'(r_h)}{2}=0
	  \end{align}
	  for $r_h$, which yields three solutions, only one of which is viable, and is given by 
	  \begin{align}
	  	r_{ext}=-\frac{3^{1/3}\Lambda+\left(9l\Lambda^2+\sqrt{3}\sqrt{\Lambda^3(-1+27l^2\Lambda)}\right)^{2/3}}{3^{2/3}\Lambda\left(9l\Lambda^2+\sqrt{3}\sqrt{\Lambda^3(-1+27l^2\Lambda)}\right)^{1/3}}.\label{eq:rext}
	  \end{align}
	  We note that demanding the surface gravity to be zero is equivalent to demanding the effective temperature to be zero as seen from Eqs.~\eqref{eq:Tsg} and \eqref{eq:temp-comparison}. The extremal radius is a function of both the minimal length and the cosmological constant which in this case is negative. An example of where this extremal radius is located can be seen in Fig.~\ref{fig:Cv} depicted as a thick vertical black line for the values $l=0.1$ and $\Lambda=-1$ with the approximate radius being $r_{ext}\approx 0.193$. This indicates that the heat capacity exhibits a divergent behavior at radii larger than the extremal limit, which makes the study of phase structure meaningful. We see that as we increase $l$ past a certain value, we have no longer infinities/discontinuities, which is a restatement of the result we obtained in Fig.~\ref{fig:PT-AdS}, i.e.,~the existence of a critical point. This behavior agrees, as required for consistency, with the critical values we derived in Eq.~\eqref{eq:crit-values}. For $\Lambda=-1$, it is easy to find that the corresponding critical minimal length is $l^{(c)}\approx 0.156$. We see in Fig.~\ref{fig:Cv} that for critical lengths above this value, no phase transition occurs since there are no discontinuities of the heat capacity.
	  
	  We now derive, once again, using this time the heat capacity, the critical values for the minimal length, and the cosmological constant. To do that we have to examine the discriminant of the denominator of the heat capacity. It is given by 
	  \begin{align}
	  	\Delta =-4 \left(33l^2\Lambda +2170l^4\Lambda^2+33561 l^6\Lambda^3+432l^8\Lambda^4\right)
	  \end{align}
	  Solving the equation $\Delta=0$, treating $\Lambda$ as a variable, yields four solutions. Two of the roots are $\Lambda=0$ and $\Lambda=\Lambda^{(c)}=-8\pi P^{(c)}$, and the two remaining ones are 
	  \begin{align}
	  	\Lambda_{1}\approx-\frac{77.6228}{l^2}, \quad \Lambda_{2}\approx -\frac{0.0402495}{l^2}.
	  \end{align}
	 The discriminant in between these solutions has a positive sign, but yields only negative or imaginary solutions for the denominator, which means that in this region we do not have any discontinuities for the heat capacity. We conclude that for $\Lambda<\Lambda^{(c)}$ or, equivalently, for $P>P^{(c)}$, no discontinuities occur, and therefore no phase transitions in agreement with what we saw/derived previously using the critical point equations.

	  \subsubsection{Critical exponents}\label{sec:crit-exp}
	  The behavior of thermodynamic quantities near the critical point is described by the critical exponents. They do not depend on the details, i.e.,~microscopic structure of the physical system, but only on some general features. To calculate them in AdS for the RBH under consideration, we first define the quantity 
	  \begin{align}
	  	t=\frac{T_{AdS}}{T^{(c)}_{AdS}}-1,
	  \end{align}
	  which is called the reduced temperature. The main relations we are going to use are the following
	  \begin{align}
	  	C_{V}\propto |t|^{-\alpha},&\quad  g=v_{l}-v_{s}\propto |t|^{\beta},\nonumber \\
	  	\kappa_{T}=-\frac{1}{V}\frac{\partial V}{\partial P}\bigg|_{T}\propto |t|^{-\gamma},&\quad |P-P^{(c)}|\propto |V-V^{(c)}|^{\delta},
	  \end{align}
	  where $C_{V}$ is the heat capacity at constant volume, $g$ is the order parameter, and $\kappa_{T}$ is the isothermal compressibility. 
	 
	  We start with the calculation of the critical exponent $\alpha$. We see from Eq.~\eqref{eq:Cv-ads} that the heat capacity is independent of the temperature, and since this is also true for the entropy as seen from Eq.~\eqref{eq:entropy}, we conclude that $\alpha=0$. 
	  
	  \begin{figure}[htbp]
	  	\hspace*{-1cm}
	  	\includegraphics[scale=0.85]{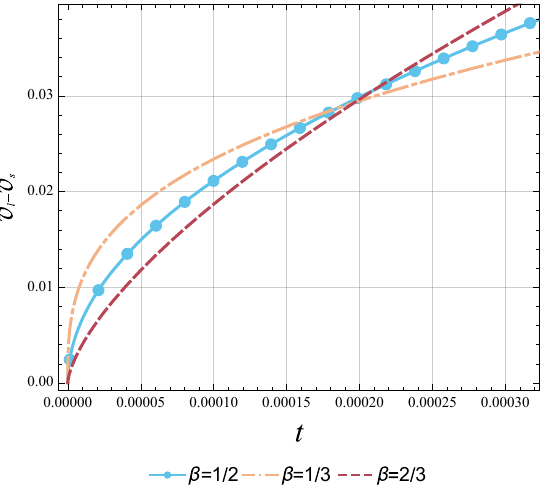}
	  	\caption{Behavior of the order parameter $g$ near the critical point as a function of the reduced temperature. Different colors/types represent different values of the coefficient $\beta$. The light blue dots represent numerical values of the points $(t,v_l-v_s)$ obtained by keeping fixed minimal length $l=0.1$ and varying the cosmological constant $\Lambda$ in the vicinity of the critical value $\Lambda^{(c)}\approx -2.445$.} 
	  	\label{fig:beta}
	  \end{figure}
	  
	  To find the critical exponent $\beta$ we need to find a way to relate the difference between the reduced volumes of the small and large black hole phases with the reduced temperature near the critical point. To proceed in an analytic way one can identify the points where the two phases coexist, i.e.,~$F|_{v_{s}}=F|_{v_{l}}$ and $T_{AdS}|_{v_s}=T_{AdS}|_{v_l}$, where $v_s$ and $v_l$ are the reduced volumes for small and large black holes, respectively. This requires finding solutions of fifth-degree polynomials to identify the point of the phase transition. Since this is not possible, we focus on finding the critical exponent $\beta$ in a numerical way. We keep the minimal length fixed at $l=0.1$ and start varying $\Lambda$ near the critical value $\Lambda_{c}=-8\pi P^{(c)}\approx -2.445$ calculated from Eq.~\eqref{eq:crit-values}. Then, we find numerical solutions for the point of the phase transition, and by identifying the temperature $T_{AdS}$ at this point, we also have the reduced temperature. For every value of the cosmological constant, we generate different points shown in Fig.~\ref{fig:beta}. By appropriate fitting of curves of the form $|t|^{\beta}$ we see that the exponent $\beta=1/2$ is a perfect fit for the points obtained numerically. 
	  
	   For the critical exponent $\gamma$ we need to plot the isothermal compressibility $\kappa_{T}$ as a function of the reduced temperature, although in this case we can do the calculation analytically. Using the monotonicity of the volume with respect to the horizon radius, we have 
	   \begin{align}
	   	\kappa_{T}=-\frac{1}{V}\left(\frac{\partial V}{\partial r_h}\right)\left(\frac{\partial P}{\partial r_H}\right)^{-1}\bigg|_{T_{AdS}}.
	   \end{align} 
	   Calculation near the critical point yields a behavior 
	   \begin{align}
	   	\kappa_{T}\propto |t|^{-1},
	   \end{align}
	   and thus we conclude that the exponent $\gamma=1$.
	   
	   The last exponent we calculate is $\delta$. This requires evaluating the absolute value of the difference between the pressure P given by Eq.~\eqref{eq:pressure} and the critical pressure $P^{(c)}$ given by Eq.~\eqref{eq:crit-values}. This can be done straightforwardly if we also use the critical radius $r^{(c)}_{h}$. Once we have the difference $|P-P^{(c)}|$, we can plot it and compare it with the difference $|V-V^{(c)}|$ near the critical point, and we see that the best-fit is for $\delta=3$.   
	   
	   In summary, the four critical exponents are given by 
	   \begin{align}
	   	\alpha=0,\quad \beta=\frac{1}{2},\quad \gamma=1,\quad \delta=3,
	   \end{align}
	   which are exactly the values expected from mean-field theory, even though we have a deviation from the critical ratio $3/8$. 
	   
	\subsection{Embedding in Minkowski}\label{sec:PT-M}
	
	\begin{figure*}[!htbp]
		\centering
		\begin{tabular}{@{\hspace*{0.0\linewidth}}p{0.45\linewidth}@{\hspace*{0.05\linewidth}}p{0.45\linewidth}@{}}
			\centering
			\subfigimg[scale=0.85]{(a)}{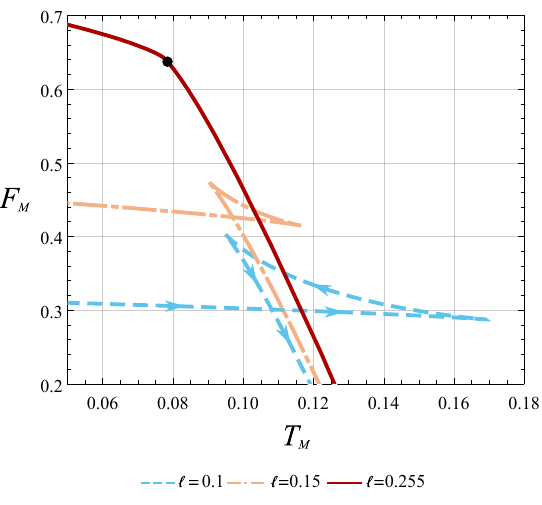} &
			\subfigimg[scale=0.85]{(b)}{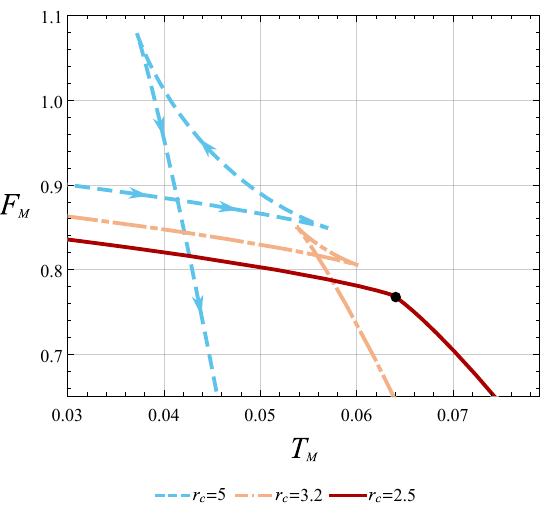}
		\end{tabular}
		\caption{(a) Free energy $F_{M}$ as a function of the temperature $T_{M}$ for a constant value of the cavity radius $r_c=2$ with varying minimal length $l$; (b) Free energy $F_{M}$ as a function of the temperature $T_{M}$ for a constant value of the minimal length $l=0.3$ with varying cavity radius $r_c$; In both cases, the arrows denote the direction in which the horizon radius $r_h$ increases. For the sake of readability, we omit the arrows in the remaining curves, although the same pattern is implied.} \label{fig:PT-M}
	\end{figure*}
	In this section, we will study the thermodynamics in Minkowski spacetime starting with the definition of the proper thermodynamic quantities and continuing with the phase structure. As we saw at the beginning of Sec.~\ref{sec:thermodynamics}, here we no longer have an effective potential that confines the radiation as in the AdS case, so the implementation of an isothermal cavity is necessary to provide thermal equilibrium for the system. The temperature of the black hole is found by Eq.~\eqref{eq:Tds} taking the limit $\Lambda\rightarrow 0$ since a cosmological constant is not present in the spacetime. We find that 
	\begin{align}
		T_{M}=\frac{r^3_c(r_h-2l)(l+r_h)^2}{4\pi (l+r_c)^3r^4_h\sqrt{\frac{-(r_c-r_h)(3l^2r_cr_h-r^2_cr^2_h+l^3(r_c+r_h))}{(l+r_c)^3r^2_h}}}.\label{eq:Tm}
	\end{align}
    Similarly, we calculate the conjugate potential for the minimal length $l$ and cavity area $A_c$ through Eq.~\eqref{eq:phi} and \eqref{eq:lambda}, respectively. They are given by 
    \begin{align}
    	\Phi_{M}=\frac{3r^3_c(r_c-r_h)(l+r_h)^2}{2(l+r_c)^4r^2_h\sqrt{\frac{-(r_c-r_h)(3l^2r_cr_h-r^2_cr^2_h+l^3(r_c+r_h))}{(l+r_c)^3r^2_h}}},
    \end{align}
    and 
    \begin{widetext}
    \begin{align}
    	\lambda_{M}=\frac{1}{48\pi r_c}\Bigg(6+\frac{3r^2_c(2l-r_c)(l+r_h)^3}{(l+r_c)^{5/2}\sqrt{-(r_c-r_h)(3l^2r_cr_h-r^2_cr^2_h+l^3(r_c+r_h))}}-6\sqrt{\frac{-(r_c-r_h)(3l^2r_cr_h-r^2_cr^2_h+l^3(r_c+r_h))}{(l+r_c)^3r^2_{h}}}\Bigg).
    \end{align}
     \end{widetext}
	Two additional quantities need to be computed: First, the internal energy of the system, which is given by setting $\Lambda=0$ in Eq.~\eqref{eq:energy}, and leads to
	\begin{align}
		E_{M}=r_c\left(1-\sqrt{1-\frac{r^2_c(l+r_h)^3}{(l+r_c)^3r^2_h}}\right),\label{eq:Em}
	\end{align}
	and second, the entropy of the system which remains the Bekenstein-Hawking entropy as seen from the general result of Eq.~\eqref{eq:entropy}. The above quantities give rise to a proper Smarr formula of the form 
	\begin{align}
		E_{M}=2T_{M}S+\Phi_{M}l+2\lambda_{M}A_{c}.
	\end{align} 
	The above quantities are the ones that will be used for the study of the phase structure, although it is worth writing the explicit form of the thermodynamic quantities when we position the cavity at infinity. To indicate the absence of the cavity, we will use the superscript $(0)$ in these expressions. The temperature is given by
	\begin{align}
		T^{(0)}_{M}=\frac{(l+r_h)^2(r_h-2l)}{4\pi r^4_h},
	\end{align}
	and it is easy to check that it is connected to the surface gravity by the relation 
	\begin{align}
		T^{(0)}_{M}=\left(1+\frac{l}{r_h}\right)^3\mathcal{T}|_{\Lambda=0},
	\end{align}
	as was the case for AdS. The conjugate potential $\Phi$ in the limit $r_c\rightarrow \infty$ takes the form 
	\begin{align}
		\Phi^{(0)}_{M}=\frac{3(l+r_h)^2}{2r^2_h},
	\end{align}
	while the energy becomes 
	\begin{align}
		E^{(0)}_{M}=\frac{(l+r_h)^3}{2r^2_h},
	\end{align}
	which is the Arnowitt-Deser-Misner mass \cite{ADM:59} of the spacetime. In terms of these quantities, it is obvious that the Smarr formula becomes 
	\begin{align}
			E^{(0)}_{M}=2T^{(0)}_{M}S+\Phi^{(0)}_{M}l,
	\end{align}
	since we do not have a term due to the cavity. Having defined the proper thermodynamic quantities $T_{M}$, $\Phi_{M}$ and $E_{M}$ we proceed with the calculation of the free energy and move on to the study of the phase structure. The free energy is given by 
	\begin{align}
		F_{M}=E_{M}-T_{M}S,
	\end{align}  
	which is calculated using Eqs.~\eqref{eq:Tm}, \eqref{eq:Em} and \eqref{eq:entropy}, to be 
	\begin{widetext}
	\begin{align}
		F_{M}=\frac{r_c}{4(r_c+l)^3}\left(4l^3+12l^2r_c+12lr^2_c+4r^3_c-\frac{4l^3+12l^2r_c+4r^3_c-\frac{3r^2_c}{r^2_h}(2l^3+5l^2r_h+r^3_h)}{\sqrt{1-\frac{r^2_c(l+r_h)^3}{(l+r_c)^3r^2_h}}}\right).
	\end{align}
	\end{widetext}
	We see from Fig.~\ref{fig:PT-M} that we have the swallowtail structure once again, and the presence of the minimal length prevents the Hawking-Page transition from occurring.  In Fig.~\ref{fig:PT-M}(a) we keep the cavity radius constant at $r_c=2$ and vary the minimal length. We see that with increasing temperature we are led to a first-order phase transition, which is from a small to a large black hole since the direction of increasing $r_h$ is the same as that of the increasing temperature (the direction of increasing horizon radius is also indicated with arrows on the dashed light blue curve in Fig.~\ref{fig:PT-M}). We have a critical value $l_{c}\approx 0.25$ of the minimal length above which no phase transition occurs. The same behavior is encountered in Fig.~\ref{fig:PT-M}(b), where the minimal length scale remains constant at $l=0.3$ and we vary the cavity radius $r_c$. Again, a first-order small to large phase transition occurs. There is a critical cavity radius $r^{(c)}_c\approx 2.5$ below which no phase transition occurs.
	
	 \begin{figure}[htbp]
		\hspace*{-0.9cm}
		\includegraphics[scale=0.85]{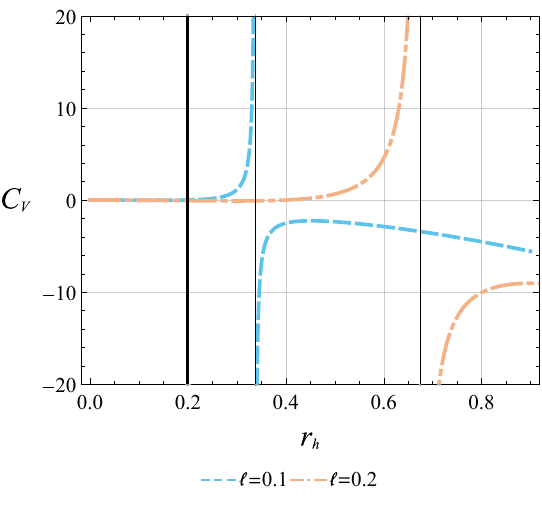}
		\caption{Heat capacity $C_{V}$ as a function of the horizon radius $r_h$ for Minkowski spacetime. Different colors/types represent different values of the minimal length parameter $l$. The thin vertical black lines correspond to infinities/discontinuities of the heat capacity, while the thick vertical black line represents the minimal radius enforced by the extremal limit for the case $l=0.1$.  } 
		\label{fig:Cv-mink}
	\end{figure}
	
	\subsubsection{Constraints on the minimal length}
	
	In Ref.~\cite{CLMMOS:23} corrections to the perihelion precession angle of test particles' orbits were found to scale linearly with the minimal length $l$. Using this result, it is possible to check if the bounds imposed on the minimal length from the requirement of thermodynamic stability and the extremal limit are compatible with available astrophysical data for the orbits of the S2 star around the Sagittarius A*. In this paper, we have an effective temperature, that differs from the one linked to surface gravity, and thus it will prove useful to once again check the minimal length's range of values. We start our analysis by taking the limit of $\Lambda\rightarrow 0^{-}$ of the heat capacity defined by Eq.~\eqref{eq:Cv-ads}. We find that 
		\begin{figure}[htbp]
		\hspace*{-0.9cm}
		\includegraphics[scale=0.85]{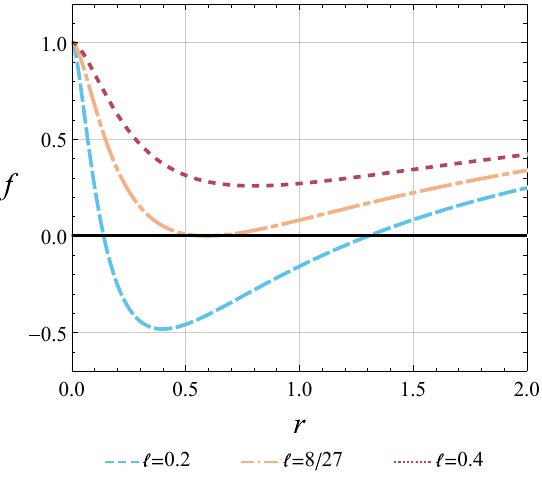}
		\caption{Metric function $f$ as a function of the radial coordinate $r$ with vanishing cosmological constant $\Lambda=0$ and mass $m=1$. Different colors/types represent different values of the minimal length.} 
		\label{fig:fmink}
	\end{figure}	
	
		\begin{figure*}[!htbp]
		\centering
		\begin{tabular}{@{\hspace*{0.0\linewidth}}p{0.45\linewidth}@{\hspace*{0.05\linewidth}}p{0.45\linewidth}@{}}
			\centering
			\subfigimg[scale=0.85]{(a)}{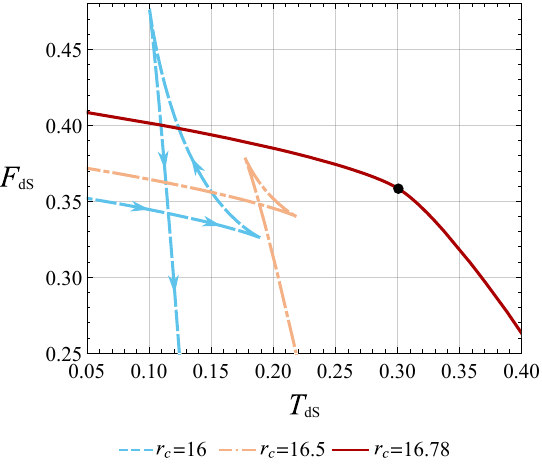} &
			\subfigimg[scale=0.85]{(b)}{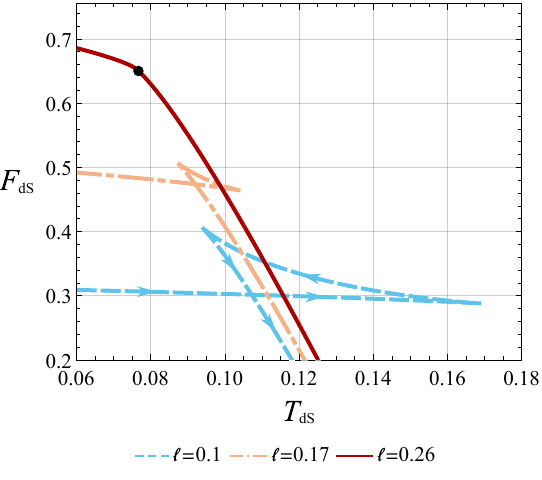}
		\end{tabular}
		\caption{(a) Free energy $F_{dS}$ as a function of the temperature $T_{dS}$ for constant value $l=0.1$ for the minimal length and constant cosmological constant $\Lambda=0.01$ with varying cavity radius $r_c$; (b) Free energy $F_{dS}$ as a function of the temperature $T_{dS}$ for constant value of cavity radius $r_c=2$ and cosmological constant $\Lambda=0.01$ with varying minimal length $l$. In both cases, the arrows denote the direction in which the horizon radius $r_h$ increases. For the sake of readability, we omit the arrows in the remaining curves, although the same pattern is implied.} \label{fig:PT-dS}
	\end{figure*}

	\begin{align}
		C_{V}=\frac{2\pi r^2_h(l+r_h)(2l-r_h)}{r^2_h-lr_h-8l^2},\label{eq:Cv-mink}
	\end{align} 
	and it is shown in Fig.~\ref{fig:Cv-mink}. We can see that there is only one infinity/discontinuity that separates thermodynamically stable and unstable regions. We have to point out that for this analysis, we have considered the cavity radius to be at infinity to retrieve the ordinary thermodynamic quantities. The point at which the discontinuity occurs is the positive root of the denominator of Eq.~\eqref{eq:Cv-mink}, which is 
	\begin{align}
		r_{disc}=\frac{1+\sqrt{33}}{2}l.
	\end{align} 
	For values of the horizon radius smaller than $r_{disc}$, we have a thermodynamically stable RBH. It will be much more intuitive and useful for comparison with observational data to impose bounds on $l$ with respect to the mass of the black hole. The extremal limit is the result of the outer horizon merging with the inner one forming a degenerate horizon \cite{MS:23}. In this case, we note that this occurs when $l$ takes the value
	\begin{align}
		\tilde{l}=\frac{8m}{27},\label{eq:l-tilde}
	\end{align}
	although we will refrain from calling a ``critical" length to avoid confusion with the one introduced while studying phase transitions.

	We can see from Fig.~\ref{fig:fmink} that the value $l=\tilde{l}$ corresponds to one double root of the function indicating the presence of an extremal black hole. For minimal lengths $l>\tilde{l}$ there is no black hole and this geometry represents a horizonless configuration whereas for $l<\tilde{l}$ we have an RBH. This condition places a constraint on the minimal length, which is a crucial factor to consider when assessing the feasibility of RBHs. Using Eq.~\eqref{eq:rext} and taking the limit $\Lambda \rightarrow 0^{-}$, we find that the extremal radius is $r_{ext}=2\tilde{l}$. This result can also be obtained directly from Eq.~\eqref{eq:Tm} by demanding the temperature to vanish. For the existence of a thermodynamically stable black hole, we must have then that $r_{disc}>r_{ext}$, which enforces the following lower bound on the minimal length: 
	\begin{align}
		r_{disc}>r_{ext}\Rightarrow  l>\frac{4}{1+\sqrt{33}}\tilde{l}. 
	\end{align}
	 Combining the constraints mentioned above we have the following bounds for the minimal length	 
	 
	\begin{align}
		\frac{4}{1+\sqrt{33}}\tilde{l}<l<\tilde{l},
	\end{align}
	which can be rewritten, using Eq.~\eqref{eq:l-tilde}, as
	\begin{align}
		0.176<\frac{l}{m}<0.296.
	\end{align}
	This demonstrates that using the effective temperature derived from the Euclidean path integral approach leads to a lower bound than the one presented in Ref.~\cite{CLMMOS:23}, hence allowing for even smaller values of the parameter $l$, yet still affirming the potential of identifying the observed ultracompact objects as RBHs. We note that the bounds derived in this subsection are calculated within the framework of an asymptotically Minkowski spacetime. It is important to emphasize that the ultracompact objects under observation are situated within a spacetime characterized by a positive cosmological constant. Consequently, there will be corrections to these bounds due to its presence but we expect them to be negligible since the cosmological constant is of the order $10^{-122}l^{-2}_{p}$ \cite{planck}. 
	
	\subsection{Embedding in de Sitter}\label{sec:PT-dS}
	
	Finally, in this section we are going to study the most physically relevant case, that of an asymptotically de Sitter spacetime. The cavity radius is now restricted to a domain between the black hole horizon and the cosmological horizon.  The proper thermodynamic quantities in this case are defined by Eqs.~\eqref{eq:Tds}, \eqref{eq:phi}, \eqref{eq:energy}, \eqref{eq:V} and \eqref{eq:lambda} but now with $\Lambda>0$. The sign of $\Lambda$ indicates that the quantity $P=-\Lambda/8\pi<0$ should be interpreted as tension and not pressure. We explicitly add to these quantities a subscript $dS$ to separate them from the general quantities derived from the Euclidean action in Sec.~\ref{sec:thermodynamics}, even though they are identical, and also to maintain consistency with the notation used in previous subsections. These thermodynamic quantities define the following Smarr relation written as 
	\begin{align}
		E_{dS}=2T_{dS}S+\Phi_{dS}l+2\lambda_{dS} A_{c}-2PV_{dS}.
	\end{align}
	The free energy $F_{dS}$ is calculated in the same manner as in previous subsections by using the mean thermal energy, the temperature, and the entropy, 
	\begin{align}
		F_{dS}=E_{dS}-T_{dS}S.
	\end{align}
	We omit the full expression for the free energy in the main text and provide it in Appendix \ref{sec:app:therm-potentials} instead due to its extensive nature.

	 \begin{figure}[htbp]
	 	\hspace{-0.5cm}
		\includegraphics[scale=0.9]{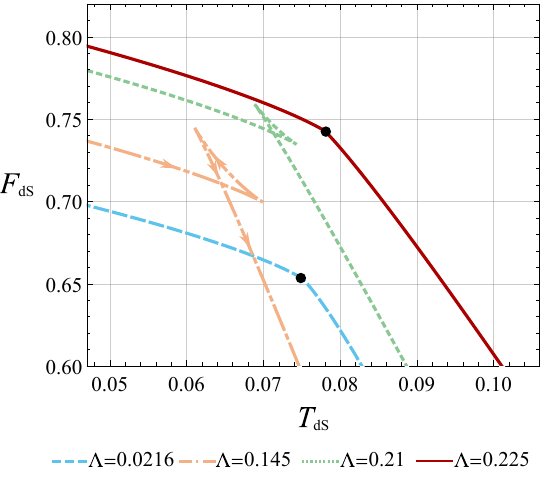}
		\caption{Free energy $F_{dS}$ as a function of the temperature $T_{dS}$ for a constant value of the minimal length $l=0.256$ and constant cavity radius $r_c=2.01$ with varying cosmological constant $\Lambda$. The arrows denote the direction in which the horizon radius $r_h$ increases. For the sake of readability, we omit the arrows in the remaining curves, although the same pattern is implied.} 
		\label{fig:PT-dS-variable-Lambda}
	\end{figure}
	
	We once again create parametric plots of the free energy as a function of the temperature using the horizon radius $r_h$ as a parameter. We see from Fig.~\ref{fig:PT-dS} that we observe the same behavior as in the asymptotically flat spacetime with the variation of the cavity radius and the minimal length leading us to the swallowtail behavior. We can see that the first-order small to large phase transitions are in general present when the cavity radius is close to the black hole horizon $r_h$ or the cosmological horizon $r_{cosm}$. Fig.~\ref{fig:PT-dS}(a) represents the case where the cavity is close to the cosmological horizon $r_{cosm}\approx \sqrt{(3/\Lambda)}\approx 17.32$. It is important to acknowledge that this approximation is because the cosmological horizon is influenced by the presence of the black hole, causing it to be situated closer to the black hole horizon than it would be if the black hole was absent. The same behavior is exhibited, as seen in Fig.~\ref{fig:PT-dS}(b), when the cavity is close to the black hole horizon and we vary the minimal length.

\begin{figure}[htbp]
	\hspace*{-1cm}
	\includegraphics[scale=0.9]{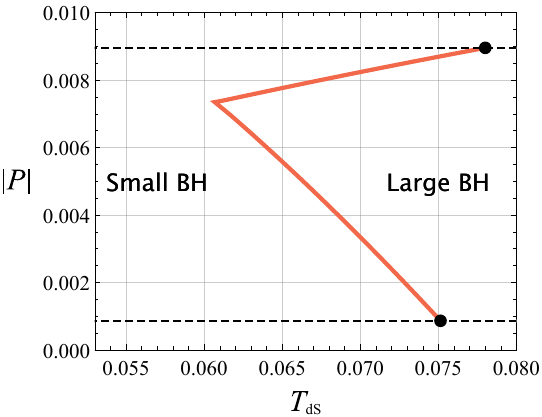}
	\caption{Coexistence line for the RBH model in de Sitter spacetime. The black dots represent second-order phase transitions. The two horizontal dashed black lines illustrate the absolute value of the two critical pressures. The bottom line corresponds to $|P^{(c)}_{1}|=|\Lambda^{(c)}_{1}|/8\pi\approx 0.00085$ and the top one to $|P^{(c)}_{2}|=|\Lambda^{(c)}_{2}|/8\pi\approx 0.0089$. } 
	\label{fig:twosecord}
\end{figure}

	However, there is a main difference in comparison with the AdS spacetime. The variation of $\Lambda$ leads to the swallowtail, but instead of terminating at one point, which represents a second-order phase transition, we now have two values of $\Lambda$ that can lead to the closure of the swallowtail, and therefore it is transformed into a swallowtube. The one obvious value is $\Lambda\rightarrow 0^{+}$ since we are considering a de Sitter spacetime and the cosmological constant must remain positive, but independently of this restriction we see from Fig.~\ref{fig:PT-dS-variable-Lambda} that for certain values of $l$ and $r_c$ the termination happens at a value of the parameter $\Lambda^{(c)}_{1}\approx 0.0216$ larger than zero.  A second value of the cosmological constant $\Lambda^{(c)}_{2}\approx 0.225$ also leads to a termination of the first-order phase transitions. Therefore in de Sitter, we have two critical points with the coexistence line shown in Fig.~\ref{fig:twosecord}.

To summarize, in a spacetime with a positive cosmological constant, under the choice of an appropriate parameter for the minimal length and the cavity radius, there is an interval $(P_{min},P_{max})$ in which first-order small to large phase transitions are possible, but outside of which no phase transitions occur, i.e.,~the small to large phase transition is compact. This behavior is in stark contrast with AdS, where only one critical point exists, and for temperatures or pressures higher than this no phase transitions occur. Direct comparison of Fig.~\ref{fig:twosecord} with Fig.~\ref{fig:coex-line} makes evident the different and richer phase structure of the RBH in de Sitter spacetime in comparison with AdS. This novel feature can be solely attributed to the existence of a cosmological horizon, as the swallowtube behavior is absent in both AdS and Minkowski spacetime, rather than being influenced by the cavity's presence \cite{SM:18}.

\section{Conclusions}\label{sec:conlusions}

Studying various RBH models is crucial for gaining insights into how the smearing of the singularity, found in mathematical black holes, influences the classical sector.  Assuming the existence of a quantum gravity theory that achieves such a regularization while preserving the hallmark feature of the horizon, RBHs emerge as the sole viable description for ultracompact objects. Numerous models have been proposed in the literature to describe such objects, with the common thread being the introduction of a minimal length scale, while allowing for distinct asymptotic behaviors. Most of these models can be described in the context of general relativity, having as a source an NED theory coupled to gravity with the incorporation of a magnetic charge, although alternative theories to describe such geometries are possible \cite{NN:23}.

In this paper we analyzed the way to generate the model proposed in Ref.~\cite{CLMMOS:23} in the framework of general relativity and derived the appropriate thermodynamic quantities using the Euclidean path integral approach. We conducted a comprehensive analysis of the thermodynamic properties and the phase structure of this model across anti-de Sitter, Minkowski and de Sitter spacetimes. Our findings imply/reveal a fundamental connection between the presence of a minimal length and the absence of the Hawking-Page transition in all three scenarios. Additionally, we observed that the method of singularity regularization influences the extent of the deviation from mean-field theory critical ratio in AdS. Both of these results arise from the fact that the singularity is not present. In AdS and Minkowski spacetime, the phase structure exhibits a typical swallowtail behavior, characterized by a unique critical point. However, in de Sitter spacetime, the swallowtail transforms into a swallowtube due to the emergence of a second critical point under specific values of the cavity radius and minimal length. In the asymptotically flat spacetime we established bounds on the minimal length, assuming the existence of a thermodynamically stable RBH. Our results indicate a lower bound smaller than the one previously predicted, implying the possibility of even smaller minimal length scales underpinning the possible existence of RBHs.

Our analysis has been restricted to a comparison of spherically symmetric RBHs, although realistic ultracompact objects are rotating and consequently characterized by axial symmetry. On this account, it will be intriguing to extend our investigation to include rotating RBHs. This extension should be based on the Euclidean path integral approach and will allow us to explore the thermodynamic properties and phase structure of these axially symmetric objects. Last but not least, the implications of singularity regularization in spherically symmetric setups have become evident in this paper, and thus analogous extensions should be considered in axially symmetric spacetimes. Furthermore, this extension to physically realistic scenarios could potentially allow us to make contact with gravitational wave observations.

	\acknowledgments
	I would like to thank Sebastian Murk, Fil Simovic, and Daniel Terno for useful discussions and helpful comments. I would also like to thank Zixin Huang for providing useful comments/advice regarding the figures created in this paper. I.S. is supported by an International Macquarie University Research Excellence Scholarship.
	
	\appendix
	
	\section{Calculation of reduced Euclidean action} \label{sec:app:Ir}
	
	In this section of the appendix we provide the explicit calculation of each term introduced in the reduced action, defined as
	 \begin{align}
		I_{r}=I_{EH}+I_{GHY}+I_{M}+I_{EMB}-I_{0}.
	\end{align} 
    We start with the Einstein-Hilbert action $I_{EH}$, given by Eq.~\eqref{eq:Ieh-initial}. We first perform the transformation $t\rightarrow -i\tau$ (Wick rotation) in the metric of Eq.~\eqref{eq:S2:metric}, where $\tau$ is the Euclidean time, to obtain the Euclidean metric,
	\begin{align}
		ds^2=f(r)d\tau^2+\frac{dr^2}{f(r)}+r^2d\Omega_{2},
	\end{align}
	and we find that the Ricci scalar is
	\begin{align}
		R=\frac{2-f''(r)r^2-4rf'(r)-2f(r)}{r^2}.\label{eq:Ricci}
	\end{align} 
	Therefore, 
	\begin{align}
		I_{EH}=-\frac{1}{16\pi}\int_{0}^{\beta_{h}}d\tau\int_{0}^{\pi}d\theta\int_{0}^{2\pi}d\phi\int_{r_h}^{r_c}r^2\sin{\theta}Rdr,
	\end{align}
	where the integration of the radial part is up to the cavity radius $r_c$ which is subject to restrictions in the presence of a cosmological horizon $r_{cosm}$ with $r_h<r_c<r_{cosm}$. If a cosmological horizon, is not present then the cavity radius is not bounded from above and we are allowed to take the cavity to be at infinity. The upper bound $\beta_{h}$ of the integral over $d\tau$ is chosen as to eliminate the conical singularity at the origin of $\tau-r$ plane in the Euclidean section, which corresponds to the black hole horizon in the Lorentzian geometry. After integration of the angular part and integration by parts of the radial part, we have 
	\begin{align}
		I_{EH}=\frac{\beta_{h}}{4}\left(2(r_c-r_h)-2r_cf(r_c)-r^2_cf'(r_c)\right)-\frac{\beta_{h}}{4}r^2_hf'(r_h).
	\end{align}
	Upon identifying the periodicity $\beta_{h}$ with the Killing surface gravity $\kappa$  \cite{FN:98,S:95} through the relation
	\begin{align}
		\beta^{-1}_{h}=\frac{\kappa}{2\pi}=\frac{f'(r_h)}{4\pi},\label{eq:periodicity}
	\end{align}
	we finally obtain 
	\begin{align}
		I_{EH}=-\pi r^2_h-\frac{\beta_h}{4}\left(2(r_c-r_h)-2r_cf(r_c)-r^2_cf'(r_c)\right),\label{eq:Ieh}
	\end{align}
	where the first term is shown to be the black hole entropy in Sec.~\ref{sec:thermodynamics}. We now proceed with the next term of the action which is the Gibbons-Hawking-York term 
	\begin{align}
		I_{GHY}=\frac{1}{8\pi}\int_{\partial M}\sqrt{k}K=\frac{1}{8\pi}\int_{0}^{\beta_{h}}d\tau \int_{0}^{\pi}d\theta \int_{0}^{2\pi}d\phi\sqrt{k}K\bigg|_{r_c}. \label{eq:GHY-initial-1}
	\end{align}
	The induced metric $k_{ab}$ on the boundary hypersurface is given by the line element,
	\begin{align}
		ds^2=f(r_c)d\tau+r^2_cd\Omega_{2},\label{eq:induced-metric}
	\end{align}
	and the square root of the determinant $k$ of the induced metric is 
	\begin{align}
		\sqrt{k}=\sqrt{f(r_c)}r^2_c\sin{\theta}.\label{eq:det-ind-metric}
	\end{align}
	The trace of the extrinsic curvature $K_{ab}$ evaluated at the boundary is 
	\begin{align}
		K(r_c)=-\frac{2\sqrt{f(r_c)}}{r_c}-\frac{f'(r_c)}{2\sqrt{f(r_c)}}.\label{eq:trace-ext-curv}
	\end{align}
	Combining now Eqs.~\eqref{eq:GHY-initial-1}, \eqref{eq:det-ind-metric} and \eqref{eq:trace-ext-curv}, we have that
	\begin{align}
		I_{GHY}=\frac{\beta_{h}}{2}r^2_c K(r_c)\sqrt{f(r_c)}.\label{eq:Ighy}
	\end{align}
	We now proceed with the calculation of the matter part of the action given by Eq.~\eqref{eq:Im-initial}. The Lagrangian depends only on the radial coordinate so we proceed with the integration of the angular and Euclidean time parts to arrive at
	\begin{align}
		I_{M}=\frac{\beta_{h}}{4}\int_{r_c}^{r_h}r^2\mathcal{L}(r)dr.
	\end{align}
	Instead of using Eq.~\eqref{eq:L(r)}, we can produce a more general result by using the original solution of the Einstein equations for $\mathcal{L}$ which is given by Eq.~\eqref{eq:L(r)-original} and perform the integration, which gives the rather simple result 
	\begin{align}
		I_{M}=\frac{\beta_{h}}{2}\left((r_c-r_h)-r_cf(r_c)\right).\label{eq:Im}
	\end{align}
	
	Since the next action term is the electromagnetic boundary term, which vanishes as explained in Sec.~\ref{sec:thermodynamics}, we proceed with the calculation of the final term $I_{0}$. We calculate this term by demanding the reduced action to vanish in the absence of a black hole. A comment is necessary here as to what condition we need to impose. Having a look at Eq.~\eqref{eq:m}, we see that the condition $r_{h}\rightarrow 0$ does not correspond to vanishing mass $m=0$. We point out that the proper condition in this case corresponds to first taking the limit $l\rightarrow 0$ and then taking the limit $r_h\rightarrow 0$. The limit cannot be taken in reverse order because there is no meaning in having a minimal length scale $l$ without the presence of a horizon radius, i.e.,~a black hole.\footnote{The existence of a nonzero minimal length scale $l$ is conceivable even in the absence of a horizon. However, under these circumstances, the object under consideration will no longer quantify as a black hole, i.e.,~a trapped region of spacetime. Instead, it would be classified as a horizonless object \cite{CRFLV:23,Bambi:book:23}. In such scenarios, the application of the Euclidean formalism, in this form, for the thermodynamic analysis becomes unsuitable, as it relies on the presence of a horizon.} Consequently, we have the following equivalence between these two conditions,
	\begin{align}
		\lim_{m\rightarrow 0} I_{r}=0 \Leftrightarrow\lim_{r_h\rightarrow 0}\left(\lim_{l\rightarrow 0}I_{r}\right)=0. \label{eq:condition}
	\end{align}  
	This condition leads to a vanishing action for the absence of a black hole, i.e.,~an empty spacetime which can either be Minkowski or AdS/dS, depending on the value of the cosmological constant. Implementing the above condition and taking into account the introduction of the new parameter $\beta$ for appropriate comparison of the spacetimes with and without a black hole, leads to the expression,
	\begin{align}
		I_{0}=\frac{1}{3}\beta (-3+r^2_c\Lambda),\label{eq:I0}
	\end{align} 
	for the subtraction term.
	
	\section{Considering $Q_{m}$ and $\sigma$ as fundamental thermodynamic variables}\label{sec:app:action-Qm}
	
	In this section of the appendix, we derive the reduced Euclidean action when our RBH is embedded in AdS spacetime, treating $Q_m$ and $\sigma$ as fundamental thermodynamic variables. Choosing the AdS case, allows us to take the cavity radius $r_c$ to infinity, simplifying considerably the expressions while still enabling a proper comparison. Using Eqs.~\eqref{eq:sigma}, \eqref{eq:Qm} and \eqref{eq:m}, we rewrite the metric function of Eq.~\eqref{eq:f(r)} solely in terms of $Q_{m}$ and $\sigma$. This results in
	\begin{align}
		f(r)=1-\frac{2^{7/4}Q^{3/2}_m r^2}{(r+2^{1/4}\sigma^{1/4}Q^{1/2}_{m})^3\sigma^{1/4}}-\frac{\Lambda}{3}r^2. \label{eq:f(r)-Qm}
	\end{align}
	
	We proceed with the calculation of the Euclidean action, using Eqs.~\eqref{eq:Ieh}, \eqref{eq:Ighy}, \eqref{eq:Im}, \eqref{eq:I0}, and Eq.~\eqref{eq:f(r)-Qm} as the metric function. This leads to the reduced action
	\begin{align}
		I_{r}=I_{r}(r_h,r_c,Q_m,\sigma,\Lambda),
	\end{align}
	where we highlight once again that $Q_{m}$ and $\sigma$ are treated as independent thermodynamic variables. Upon taking the cavity to infinity we have that
	\begin{align}
		I_{r}(r_h,Q_{m},\sigma, \Lambda)=\lim_{r_c\rightarrow \infty}I_{r}(r_h,r_c,Q_{m},\sigma, \Lambda),
	\end{align}
		and we obtain
    \begin{widetext}
    	\begin{align}
    			I_{r}(r_h,Q_{m},\sigma, \Lambda)=-\pi r^2_{h}+\beta \left(\left(\frac{r_h}{2}+\frac{2^{7/4}Q^{3/2}_{m}}{4\sigma^{1/4}}\right)-\frac{\Lambda}{6}\left(r^3_h+\frac{3\cdot 2^{3/4}Q^{3/2}_{m}}{\Lambda \sigma^{1/4}}\right)+\frac{Q^2_{m}\left(6 r^2_{h}+3\cdot 2^{5/4}Q^{1/2}_{m}r_h\sigma^{1/4}+2^{3/2}Q_{m}\sigma^{1/2}\right)}{(r_h+2^{1/4}Q^{1/2}_{m}\sigma^{1/4})^{3}}\right). \label{eq:Ir-Qm}
    	\end{align}
    \end{widetext}	
	
	Not surprisingly this action is precisely the one derived in Eq.~\eqref{eq:Ir} if we take the limit $r_c\rightarrow \infty$ and we also substitute $Q_{m}$, $\sigma$, and $m$ in terms of $r_h$ and $l$ using Eqs.~\eqref{eq:sigma}, \eqref{eq:Qm}, and subsequently Eq.~\eqref{eq:m}. Importantly, both derivations of the action of Eq.~\eqref{eq:Ir} and Eq.~\eqref{eq:Ir-Qm} are on shell, as indicated by Eq.~\eqref{eq:periodicity}, where we have selected the inverse surface gravity as the periodicity and therefore we have eliminated the conical singularity \cite{FN:98,S:95}. 
	
	However, there is a major difference between these two calculations arising from the choice of thermodynamic variables. Initially, we calculate the temperature by extremizing the action of Eq.~\eqref{eq:Ir-Qm} with respect to the horizon radius, i.e., 
	\begin{align}
	  \frac{\partial I_{r}(r_h,Q_m,\sigma, \Lambda)}{\partial r_h}=0,
	\end{align} 
	and solve for $\beta$ which leads to the following relation for the temperature:
	\begin{align}
		\mathcal{T}=\frac{1-r^2_h\Lambda}{4\pi r_h}-\frac{12 Q^2_{m}r_h}{4\pi (r_h+2^{1/4}Q^{1/2}_m\sigma^{1/4})^4}.
	\end{align}
	After using Eqs.~\eqref{eq:sigma}, \eqref{eq:Qm}, and \eqref{eq:m} in the above expression we retrieve exactly the relation \eqref{eq:Tsg}, i.e., the temperature being the surface gravity. Since now $Q_{m}$ and $\sigma$ are independent variables, they will have their own conjugate quantities in the first law and Smarr formula. We calculate them using a similar methodology as in Sec.~\ref{sec:thermodynamics}, but we only present here the expressions after we have substituted $Q_{m}$ and $\sigma$ in terms of $r_h$ and $l$. These quantities, for the magnetic charge $Q_m$, the parameter $\sigma$, and the cosmological constant $\Lambda$, respectively, are given by the following relations:
	\begin{align}
		\Psi_{H}=\frac{1}{\beta}\frac{\partial I_{r}(r_h,Q_{m},\sigma,\Lambda)}{\partial Q_{m}}, \quad K_{\sigma}=\frac{1}{\beta}\frac{\partial I_{r}(r_h,Q_m,\sigma,\Lambda)}{\partial \sigma}
	\end{align}
	and 
	\begin{align}
		\mathcal{V}=-\frac{8\pi}{\beta} \frac{\partial I_{r}(r_h,Q_{m},\sigma,\Lambda)}{\partial \Lambda},
	\end{align}  
	where now $\beta=\mathcal{T}^{-1}$. Explicit calculation yields 
    \begin{align}
    	\Psi_{H}=\frac{\sqrt{3l(l+r_h)(3-r^2_h\Lambda)}}{2 r_h(l+r_h)^3}\left(4r^3_h+6r^2_hl +4r_h l^2+l^3\right),
    \end{align}
    \begin{align}
    	K_{\sigma}=-\frac{(l+r_h)^2(l^2+4lr_h+6r^2_h)(3-r^2_h\Lambda)^2}{144lr^4_h},
    \end{align}
    \begin{align}
    	\mathcal{V}=\frac{4\pi r^3_h}{3}.
    \end{align}
	Finally, we calculate the internal energy and entropy from Eq.~\eqref{eq:E-dIr} and Eq.~\eqref{eq:S-dIr}, respectively, and we find 
	\begin{align}
		E=\frac{\partial I_{r}(r_h,Q_m,\sigma,\Lambda)}{\partial \beta}=\frac{(l+r_h)^3(3-r^2_h\Lambda)}{6r^2_h}=m,
	\end{align}
	\begin{align}
		 S=\beta \frac{\partial I_{r}(r_h,Q_m,\sigma,\Lambda)}{\partial \beta}-I_{r}(r_h,Q_m,\sigma,\Lambda)=\pi r^2_h.
	\end{align}
    The  application of the Euclidean path integral formalism and its on shell consideration leads us to the Smarr formula,
	\begin{align}
		E=2\mathcal{T}S+\Psi_{H}Q_{m}+2K_{\sigma}\sigma-2\mathcal{V}P.
	\end{align}
	We note that this is exactly the Smarr formula obtained from Hamiltonian methods as can be verified from the methodology of Ref.~\cite{ZG:18}. The choice of $Q_{m}$ and $\sigma$ as thermodynamic variables is based on the dependence of the NED Lagrangian of Eq.~\eqref{eq:Lt(F)} on $\mathcal{F}$ and therefore $Q_m$ while at the same time there is dependence on the parameter $\sigma$, which is one of the parameters $\beta_{i}$ appearing in Eq.~\eqref{eq:1st law-gen}. (We refer the reader to Ref.~\cite{ZG:18} for a discussion and a general derivation of the above using Hamiltonian methods.) 
	
	The first law takes the form
	\begin{align}
		dE=\mathcal{T}dS+\Psi_{H}dQ_{m}+K_{\sigma}d\sigma+\mathcal{V}dP,\label{eq:1stlaw}
	\end{align} 
	with $E=m$. In this derivation, the temperature coincides with the surface gravity. However, it is important to emphasize two key points: Firstly, considering $Q_{m}$ and $\sigma$ as independent variables is nonphysical, given that both depend on $m$ and $l$. Any variations of $Q_{m}$ and $\sigma$ in the first law would result in variations of $m$ and $l$ on the right-hand side of Eq.~\eqref{eq:1stlaw}. This observation leads us to the second point, emphasizing that variation of the total energy of the spacetime should be present only on the left-hand side of Eq.~\eqref{eq:1stlaw}. Therefore, it is much more physically meaningful to consider the minimal length as a fundamental parameter and allow for its variation. With this in mind, we have that
	\begin{align}
		dQ_{m}=\frac{\partial Q_{m}}{\partial m}dm+\frac{\partial Q_{m}}{\partial l}dl,
	\end{align}
	and 
	\begin{align}
		d\sigma=\frac{\partial \sigma}{\partial m}dm+\frac{\partial \sigma}{\partial l}dl.
	\end{align}
	So implementing the above in the first law of Eq.~\eqref{eq:1stlaw} leads to
	\begin{widetext}
	\begin{align}
		dE=\frac{\mathcal{T}}{(1-\Psi_{H}\frac{\partial Q_{m}}{\partial m}-K_{\sigma}\frac{\partial \sigma}{\partial m})}dS+\frac{\Psi_{H}\frac{\partial Q_{m}}{\partial l}+K_{\sigma} \frac{\partial \sigma}{\partial l}}{(1-\Psi_{H}\frac{\partial Q_{m}}{\partial m}-K_{\sigma}\frac{\partial \sigma}{\partial l})}dl+\frac{\mathcal{V}}{(1-\Psi_{H}\frac{\partial Q_{m}}{\partial m}-K_{\sigma}\frac{\partial \sigma}{\partial l})}dP,
	\end{align}
	\end{widetext}
    and consequently to the effective thermodynamic quantities
	\begin{align}
		T=\frac{\mathcal{T}}{(1-\Psi_{H}\frac{\partial Q_{m}}{\partial m}-K_{\sigma}\frac{\partial \sigma}{\partial m})},
	\end{align}
	and 
	\begin{align}
		\Phi=\frac{\Psi_{H}\frac{\partial Q_{m}}{\partial l}+K_{\sigma} \frac{\partial \sigma}{\partial l}}{(1-\Psi_{H}\frac{\partial Q_{m}}{\partial m}-K_{\sigma}\frac{\partial \sigma}{\partial l})},\quad V=\frac{\mathcal{V}}{(1-\Psi_{H}\frac{\partial Q_{m}}{\partial m}-K_{\sigma}\frac{\partial \sigma}{\partial l})}.
	\end{align}
	
	The above quantities are the conjugate potentials when we treat $l$ as a fundamental variable. If we do their explicit calculation, using the relations \eqref{eq:sigma}, \eqref{eq:Qm}, and \eqref{eq:m}, we retrieve exactly the same conjugate potentials as the ones derived by the on shell Euclidean action method in Sec.~\ref{sec:PT-AdS} and specifically Eqs.~\eqref{eq:Tads}, \eqref{eq:phi-ads}, and \eqref{eq:therm-vol-ads}. This demonstrates consistency between the two derivations.

	\section{Thermodynamic potentials}\label{sec:app:therm-potentials}
	In this part of the appendix we give the rather long expressions for the thermodynamic volume $V$, the conjugate potential for the cavity area $\lambda$ and the free energy $F$ which will have the same expression as $F_{dS}$ but with positive cosmological constant. They are calculated to be
	  \begin{widetext}
		\begin{align}
			V&=\frac{-8\pi r^3_c}{6(l+r_c)^3r^2_h\sqrt{3-r^2_c\Lambda}\mathcal{Y}}\bigg(18l^2r_cr^2_h-3r^2_h(-l^3+3l^2r_h+3lr^2_h+r^3_h)-6lr^2_hr^4_h\Lambda-2r^5_cr^2_h\Lambda-6r^3_cr^2_h(-1+l^2\Lambda)+\nonumber \\     	
			&+r^2_c[-3(l^3+3l^2r_h-3lr^2_h+r^3_h)+2r^3_h(3l^2+3lr_h+r^2_h)\Lambda]+(-2l^3r^2_h-6l^2r_cr^2_h-6lr^2_cr^2_h-2r^3_cr^2_h)\sqrt{3-r^2_c\Lambda}\mathcal{Y}\bigg),\label{eq:V}
		\end{align}

		\begin{align}
			\lambda&=\frac{1}{48\pi r_c}\Bigg(6(1-r^2_c\Lambda)+\frac{2r^2_c\Lambda}{\sqrt{3-r^2_c\Lambda}}\mathcal{X}-2\sqrt{3-r^2_c\Lambda}\mathcal{X}+\nonumber \\   	
			&+\frac{r^2_c\sqrt{3-r^2_c\Lambda}\big(3(2l-r_c)(l+r_h)^3+r^2_h(l^3(9r_c-6r_h)+3l^2(4r^2_c+r_cr_h-2r^2_h)+l(8r^3_c+3r_cr^2_h-2r^3_h)+r_c(2r^3_c+r^3_h))\Lambda\big)}{(l+r_c)^{5/2}r_h\sqrt{-(r_c-r_h)(3l^3(r_c+r_h)+3lr^2_cr^2_h(r_c+r_h)\Lambda+3l^2r_cr_h(3+r_cr_h\Lambda)+r^2_cr^2_h(-3+(r^2_c+r_cr_h+r^2_h)\Lambda))}}\Bigg),\label{eq:lambda}
		\end{align}

		\begin{align*}
			F_{dS}=\frac{-r_c}{12 r^2_h(l+r_c)^3\mathcal{Y}}\bigg(\big(12l^3r^2_h-18l^3r^2_c-45l^2r^2_cr_h+36l^2r_cr^2_h+12r^3_cr^2_h-9r^2_cr^3_h-12l^2r^3_cr^2_h\Lambda-12lr^4_cr^2_h\Lambda-4r^5_cr^2_h\Lambda+
		\end{align*}
		\begin{align}
			+9l^2r^2_cr^3_h\Lambda+6lr^2_cr^4_h\Lambda+ r^2_cr^5_h\Lambda\big)\omega+(-12l^3r^2_h-36l^2r_cr^2_h-36lr^2_cr^2_h-12r^3_cr^2_h+4l^3r^2_cr^2_h\Lambda+12l^2r^3_cr^2_h\Lambda+12lr^4_cr^2_h\Lambda+4r^5_cr^2_h\Lambda)\mathcal{Y}\bigg),
		\end{align}
	\end{widetext}
		where we have defined for simplicity 
	\begin{align}
		\mathcal{Y}=\sqrt{3-r^2_c\Lambda+\frac{r^2_c(l+r_h)^3(-3+r^2_h\Lambda)}{(l+r_c)^3r^2_h}},
	\end{align}
	\begin{align}
		\omega=\sqrt{3-r^2_c\Lambda}.
	\end{align}
	It is worth pointing out that these are the general expressions we derive from the Euclidean action and they coincide with the de Sitter case quantities.
	
	\section{Thermodynamics of the Bardeen RBH}\label{sec:app:Bardeen}
	In this part of the appendix we will derive the temperature for the Bardeen black hole model using exactly the same methodology as in Sec.~\ref{sec:thermodynamics}. This model is desrcibed by the metric function given by 
	\begin{align}
		f(r)=1-\frac{2mr^2}{(r^2+l^2)^{3/2}}-\frac{\Lambda}{3}r^2.
	\end{align}
	We can easily derive the total reduced action by substituting this metric function to the general relations \eqref{eq:Ieh}, \eqref{eq:Ighy}, \eqref{eq:Ieh} and \eqref{eq:I0} derived in Appendix~\ref{sec:app:Ir}. This metric is also derived exclusively by magnetic charge as shown in Ref.~\cite{FW:16} and the only nonvanishing components of the electromagnetic tensor are $F_{23}=-F_{32}$. Since we want to study thermodynamics in the canonical ensemble, we need to introduce an electromagnetic boundary term in the action, which will vanish once again due to the integration on a constant time slice. Combining all of the above, we arrive at the following reduced action $I_{B}$ for the Bardeen model:
	\begin{widetext}
	\begin{align}
		I_{B}=-\pi r^2_h+\beta r_c\left(1-\frac{r^2_c\Lambda}{3}-\sqrt{3-r^2_c\Lambda}\sqrt{3-r^2_c\Lambda +\frac{r^2_c(l^2+r^2_h)^{3/2}(-3+r^2_h\Lambda)}{(l^2+r^2_c)^{3/2}r^2_h}}\right).
	\end{align}
	\end{widetext}
	To find the temperature, we extremize it with respect to the horizon radius $r_h$ and solve for $\beta$. We find that the temperature given by $T=\beta^{-1}$ is calculated to be
	\begin{align}
		T=\frac{r^3_c\sqrt{l^2+r^2_h}\sqrt{3-r^2_c\Lambda}(2l^2-r^2_h+r^4_h\Lambda)}{4\pi (l^2+r^2_h)^{3/2}r^4_h\mathcal{X}_B},
	\end{align}
	where $\mathcal{X}_{B}$ has been introduced for convenience and is given by
	\begin{widetext}
	\begin{align}
		\mathcal{X}_{B}=\sqrt{\frac{l^4r^2_h(3-r^2_c\Lambda)+r^2_cr^2_h\left(3r^2_c-r^4_c\Lambda+\mathcal{Y}_{B}(-3+r^2_h\Lambda)\right)+l^2r^2_c\left(r^2_h(6-2r^2_c\Lambda+\mathcal{Y}_{B}\Lambda)-3\mathcal{Y}_{B}\right)}{(l^2+r^2_c)^2r^2_h}},
	\end{align}
	\end{widetext}
	with $\mathcal{Y}_{B}=\sqrt{l^2+r^2_h}\sqrt{l^2+r^2_c}$. We introduced this section in the appendix because we are interested in the AdS case and the critical ratio. As we explained in the main part of this paper in the AdS case we are allowed to take the cavity radius at infinity and thus we have that the temperature is given by 
	\begin{align}
		T_{B}=\frac{\sqrt{l^2+r^2_h}(-2l^2+r^2_h-r^4_h\Lambda)}{4\pi r^4_h}.
	\end{align}
	Identifying the pressure as $P=-\Lambda/8\pi$ we can solve the above equation with respect to $\Lambda$ and then find the pressure which will yield a Van der Walls-like equation of the form 
	\begin{align}
		P_{B}=\frac{T_{B}}{2\sqrt{r^2_h+l^2}}-\frac{1}{8\pi r^2_h}+\frac{l^2}{4\pi r^4_h}.\label{eq:Pbardeen}
	\end{align}
	Using the above equation for the pressure, we can identify the critical radius and temperature by solving the system of equations 
	\begin{align}
		\frac{\partial P_{B}}{\partial r_h}=0,\quad \frac{\partial^2 P_{B}}{\partial r^2_h}=0
	\end{align}
	which remarkably can be solved exactly and leads to the following solutions
	\begin{align}
		r^{(c)}_{h}=\sqrt{2(2+\sqrt{10})}l,\label{eq:rcrit-bardeen}
	\end{align}
	and 
	\begin{align}
		T^{(c)}_{B}=\frac{25(31+13\sqrt{10})}{432\pi (5+2\sqrt{10})^{3/2}l}.\label{eq:Tcrit-bardeen}
	\end{align}
	The reduced volume is derived by observing the inverse coefficient of the temperature $T_{B}$ in Eq.~\eqref{eq:Pbardeen} for the pressure and it is given by 
	\begin{align}
		v_{B}=2\sqrt{r^2_h+l^2},
	\end{align} 
	with the critical value calculated, using Eq.~\eqref{eq:rcrit-bardeen}, to be
	\begin{align}
		v^{(c)}_{B}=2\sqrt{5+2\sqrt{10}}l.\label{eq:Vred-crit-bardeen}
	\end{align}
	The last quantity we need for the calculation of the critical ratio is the critical pressure which is found by substituting Eqs.~\eqref{eq:rcrit-bardeen} and \eqref{eq:Tcrit-bardeen} in \eqref{eq:Pbardeen}, and leads to
	\begin{align}
		P^{(c)}_{B}=\frac{5(53+17\sqrt{10})}{24(30+9\sqrt{10})^2\pi l^2}.\label{eq:Pcrit-bardeen}
	\end{align}
	Combining now Eqs.~\eqref{eq:Pcrit-bardeen}, \eqref{eq:Vred-crit-bardeen} and \eqref{eq:Tcrit-bardeen} we have that the critical ratio is given by
	\begin{align}
		\frac{P^{(c)}_{B}v^{(c)}_{B}}{T^{(c)}_{B}}=\frac{2}{5}=0.4.
	\end{align}


\begin{thebibliography}{100}
		
		\bibitem{GMBTK:00} A. M. Ghez, M. Morris, E. E. Becklin, A. Tanner, and T. Kremenek,
		{\href{https://doi.org/10.1038/35030032}{Nature (London) \textbf{407}, 349 (2000).}}
		
		\bibitem{Sall:02}  F. Nogueras-Lara, R. Schödel, A.T. Gallego-Calvente \textit{et al.},
		\href{https://doi.org/10.1038/s41550-019-0967-9}{Nat. Astron. \textbf{4}, 377 (2020).}
		
		\bibitem{LIGO:21} LIGO Scientific and Virgo Collaboration,
		{\href{https://doi.org/10.3847/2041-8213/abe949}{{Astrophys.\ J.\ Lett.} \textbf{913}, L7 (2021)}}.
		
		\bibitem{B:17} C. Bambi,
		{\href{https://doi.org/10.1103/RevModPhys.89.025001}{{ 
					Rev. Mod. Phys.} \textbf{89}, 025001 (2017).}}
		
		\bibitem{EHT:19} Event Horizon Telescope Collaboration,
		{\href{https://doi.org/10.3847/2041-8213/ab0e85}{{Astrophys.\ J.\ Lett.} \textbf{875}, L4 (2019)}}.
		
		\bibitem{IM:19} A. R. Ingram and S. E. Motta, 
		{\href{https://doi.org/10.1016/j.newar.2020.101524}{{New Astron. Rev.} \textbf{85}, 101524 (2019).}}
		
		\bibitem{RN:03} C. S. Reynolds and M. A. Nowak, 
		{\href{https://doi.org/10.1016/S0370-1573(02)00584-7}{{Phys. Rep.} \textbf{377}, 389 (2003).}}
		
		\bibitem{CP:19} V.\ Cardoso and P.\ Pani,
		{\href{https://doi.org/10.1007/s41114-019-0020-4}{ {Living Rev.\ Relativity} \textbf{22}, 4  (2019)}}.
		
		\bibitem{BCNS:19} L.\ Barack, V.\ Cardoso, S.\ Nissanke, and T.\ P.\ Sotiriou,
		{\href{https://doi.org/10.1088/1361-6382/ab0587}{{Classical Quantum Gravity} \textbf{36}, 143001 (2019)}}.
		
		\bibitem{M:23} S.\ Murk,
		\href{https://doi.org/10.1142/S0218271823420129}{Int.\ J.\ Mod.\ Phys.\ D \textbf{32}, 2342012 (2023)}.
		
		\bibitem{B:68} J.\ M.\ Bardeen 
		in \textit{Proceedings of the International Conference GR5} (Tbilisi University Press, Tbilisi, 1968). 
		
		\bibitem{D:92} I.\ Dymnikova, 
		\href{https://doi.org/10.1007/BF00760226}{Gen.\ Relativ.\ Gravit.\ \textbf{24}, 235 (1992)}.
		
		\bibitem{H:06} S.\ A.\ Hayward,
		\href{https://doi.org/10.1103/PhysRevLett.96.031103}{Phys.\ Rev.\ Lett.\ \textbf{96}, 031103 (2006)}.
				
		\bibitem{MM:04} P.\ O.\ Mazur and E.\ Mottola,
		\href{https://doi.org/10.1073/pnas.0402717101}{Proc.\ Natl.\ Acad.\ Sci.\ U.S.A. \textbf{101}, 9545 (2004)}.
		
		\bibitem{MM:23} P.\ O.\ Mazur and E.\ Mottola,
		\href{https://doi.org/10.3390/universe9020088}{Universe \textbf{9}, 88 (2023)}.
		
		\bibitem{E:73} H.\ G.\ Ellis,
		\href{https://doi.org/10.1063/1.1666161}{J.\ Math.\ Phys. (N.Y.)\ \textbf{14}, 104 (1973)}.
		
		\bibitem{MT:88} M.\ S.\ Morris and K.\ S.\ Thorne,
		\href{https://doi.org/10.1119/1.15620}{Am.\ J.\ Phys.\ \textbf{56}, 395 (1988)}.
		
		\bibitem{SV:19} A.\ Simpson and M.\ Visser,
		\href{https://doi.org/10.1088/1475-7516/2019/02/042}{J.\ Cosmol.\ Astropart.\ Phys.\ 02 (2019) 042}.
		
		\bibitem{LM:02} O.\ Lunin and S.\ D.\ Mathur,
		\href{https://doi.org/10.1016/S0550-3213(01)00620-4}{Nucl.\ Phys.\ \textbf{B623}, 342, (2002)}.
		
		\bibitem{M:05} S.\ D.\ Mathur,
		\href{https://doi.org/10.1002/prop.200410203}{Fortschr.\ Phys.\  \textbf{53}, 793 (2005)}.
		
		\bibitem{V:14} M.\ Visser, 
		\href{https://doi.org/10.1103/PhysRevD.90.127502}{Phys.\ Rev.\ D \textbf{90}, 127502 (2014)}.
		
		\bibitem{P:65} R.\ Penrose,
		\href{https://doi.org/10.1103/PhysRevLett.14.57}{Phys.\ Rev.\ Lett.\ \textbf{14}, 57 (1965)}.
		
		\bibitem{CRFLV:23} R. Carballo-Rubio, F. Di Filippo, S. Liberati, and M. Visser,
		\href{https://doi.org/10.48550/arXiv.2302.00028}{arXiv:2302.00028 (2023)}.
								
		\bibitem{BI:34} M.\ Born and L. Infeld,
		{\href{https://doi.org/10.1098/rspa.1934.0059}{Proc. R. Soc. A \textbf{144}, 425 (1934)}}.
		
		\bibitem{B:01} K. A. Bronnikov,
		{\href{https://doi.org/10.1103/PhysRevD.63.044005}{Phys. Rev. D \textbf{63}, 044005 (2001).}}
		
		\bibitem{ABG:00} E. Ayon-Beato and A. Garcia, 
		{\href{https://doi.org/10.1016/S0370-2693%2800%2901125-4}{Phys. Lett. B \textbf{493}, 149 (2000).}}
		
		\bibitem{FW:16} Z.-Y. Fan and X. Wang,
		{\href{https://doi.org/10.1103/PhysRevD.94.124027}{Phys. Rev. D \textbf{94}, 124027  (2016)}}.				
		
		\bibitem{balestra2008} S. Balestra  \textit{et al.},
		{\href{https://doi.org/10.1140/epjc/s10052-008-0597-3}{ Eur. Phys. J. C \textbf{55}, 57 (2008)}}.
		
		\bibitem{themacrocollaboration2002} M. Ambrosio \textit{et al.} (The MACRO Collaborations), 
		{\href{https://doi.org/10.1140/epjc/s2002-01046-9}{Eur. Phys. J. C \textbf{25}, 511 (2002).}}
		
		\bibitem{AAB:22} B. Acharya, J. Alexandre, P. Benes \textit{et al.}, 
		{\href{https://doi.org/10.1038/s41586-021-04298-1}{Nature (London) \textbf{602}, 63 (2022).}}
		
		\bibitem{BCH:73} J.\ M.\ Bardeen, B.\ Carter, and S.\ W.\ Hawking,
		{\href{https://doi.org/10.1007/BF01645742}{Commun.\ Math.\ Phys.\ \textbf{31}, 161 (1973)}}.
		
		\bibitem{CLMMOS:23} M. Cadoni,  M. De Laurentis, I. De Martino, R. D. Monica, M. Oi and A. P. Sanna,
		{\href{https://doi.org/10.1103/PhysRevD.107.044038}{Phys. Rev. D \textbf{107}, 044038 (2023)}}.
		
		\bibitem{SS:23} F. Simovic and I. Soranidis,
		{\href{https://doi.org/10.1103/PhysRevD.109.044029}{Phys. Rev. D \textbf{109}, 044029 (2024)}}.
		
		\bibitem{S:73}   L. Smarr,
		{\href{https://doi.org/10.1103/PhysRevLett.30.71}{Phys. Rev. Lett. \textbf{30}, 71 (1973)}}.
		
		\bibitem{Bambi:book:23} C.\ Bambi (ed.),
		\href{https://doi.org/10.1007/978-981-99-1596-5}{\textit{Regular Black Holes: Towards a New Paradigm of Gravitational Collapse} (Springer Singapore, 2023)}.
		
		\bibitem{BW:22} K. A. Bronnikov and R. K. Walia,
		\href{https://doi.org/10.1103/PhysRevD.105.044039}{Phys. Rev. D \textbf{105}, 044039  (2022)}
		\bibitem{RS:23} M. E. Rodrigues and M. V. de S. Silva,
		\href{https://doi.org/10.1103/PhysRevD.107.044064}{Phys. Rev. D \textbf{107}, 044064 (2023)}.
		
		\bibitem{K:63} A.\ Komar,
		{\href{https://doi.org/10.1103/PhysRev.129.1873}{Phys. Rev. \textbf{129}, 1873 (1963)}}.
		
		\bibitem{BBM:02} V. Balasubramanian, J. de Boer, and D. Minic, 
		{\href{https://doi.org/10.1103/PhysRevD.65.123508}{Phys. Rev. D \textbf{65}, 123508 (2002).}}
		
		\bibitem{B:02} R. Bousso, 
		{\href{https://doi.org/10.1103/RevModPhys.74.825}{Rev. Mod. Phys. \textbf{74}, 825 (2002).}}
		
		\bibitem{ANS:11} D. Anninos, G. S. Ng, and A. Strominger,
		{\href{https://doi.org/10.1088/0264-9381/28/17/175019}{Classical Quantum Gravity \textbf{28}, 175019 (2011).}}
		
		\bibitem{ZG:18} Y. Zhang and S. Gao,
		\href{https://doi.org/10.1088/1361-6382/aac9d4}{Classical Quantum Gravity \textbf{35} 145007 (2018)}.
		
		\bibitem{GH:77} G. W. Gibbons and S. W. Hawking,
		{\href{https://doi.org/10.1103/PhysRevD.15.2752}{Phys. Rev. D \textbf{15}, 2752 (1977)}}.
		
		\bibitem{Y:86} J. W. York, Jr.,
		{\href{https://doi.org/10.1103/PhysRevD.33.2092}{Phys. Rev. D \textbf{33}, 2092  (1986)}}.
		
		\bibitem{BBWY:90} H. W. Braden, J. D. Brown, B. F. Whiting, and J. W. York, Jr., 
		{\href{https://doi.org/10.1103/PhysRevD.42.3376}{Phys. Rev. D \textbf{42}, 3376 (1990).}} 
		
		\bibitem{MK:21} H. E. Moumni and J. Khalloufi,
		{\href{https://doi.org/10.1016/j.nuclphysb.2021.115593}{Nucl. Phys. \textbf{B973} 115593 (2021)}}.
		
		\bibitem{MK:22} H. E. Moumni and J. Khalloufi,
		{\href{https://doi.org/10.1016/j.nuclphysb.2022.115731}{Nucl. Phys. \textbf{B977} 115731 (2022)}}.
		
		\bibitem{CV:03} S. Carlip and S. Vaidya,
		{\href{https://doi.org/10.1088/0264-9381/20/16/319}{Classical Quantum Gravity \textbf{20}, 3827 (2003)}}.				
		
		\bibitem{KMT:17} D. Kubizňák, R. B. Mann and M. Teo,
		{\href{https://doi.org/10.1088/1361-6382/aa5c69}{Classical Quantum Gravity \textbf{34}, 063001  (2017)}}.
		
		\bibitem{KRT:09} D.\ Kastor, S.\ Ray, and J.\ Traschen,
		{\href{https://doi.org/10.1088/0264-9381/26/19/195011}{Classical Quantum Gravity \textbf{26}, 195011  (2009)}}.
		
		\bibitem{W:93} R.\ M.\ Wald,
		{\href{https://doi.org/10.1103/PhysRevD.48.R3427}{Phys.\ Rev.\ D \textbf{48}, R3427(R) (1993)}}.
		
		\bibitem{IW:94} V.\ Iyer and R.\ M.\ Wald,
		{\href{https://doi.org/10.1103/PhysRevD.50.846}{Phys.\ Rev.\ D \textbf{50}, 846 (1994)}}.
		
		\bibitem{R:03} D.\ A.\ Rasheed,
		{\href{https://doi.org/10.48550/arXiv.hep-th/9702087}{arXiv:hep-th/9702087 (2003)}}.
		
		\bibitem{GKK:96} G.\ Gibbons, R.\ Kallosh, and B.\ Kol,
		{\href{ttps://doi.org/10.1103/PhysRevLett.77.4992}{Phys. Rev. Lett. \textbf{77}, 4992 (1996)}}.
		
		\bibitem{CM:95} J.\ D.\ E.\ Creighton and R.\ B.\ Mann,
		{\href{https://doi.org/10.1103/PhysRevD.52.4569}{Phys. Rev. D \textbf{52}, 4569  (1995)}}.
		
		\bibitem{B:05} N.\ Breton,
		{\href{https://doi.org/10.1007/s10714-005-0051-x}{Gen. Relativ. Gravit. \textbf{37} 643–650 (2005)}}.
		
		\bibitem{BCSV:19} I.\ Booth, B.\ Creelman, J.\ Santiago, and M.\ Visser,
		{\href{https://doi.org/10.1088/1475-7516/2019/09/067}{J. Cosmol. Astropart. Phys. 09 (2019) 067.}}

		\bibitem{G:95} L.\ J.\ Garay,
		{\href{https://doi.org/10.1142/S0217751X95000085}{Int. J. Mod. Phys. A \textbf{10}, 145 (1995).}}
		
		\bibitem{BBCRG:21}  C.\ Barceló, V.\ Boyanov, R.\ Carballo-Rubio, and L.\ J.\ Garay,
		{\href{https://doi.org/10.1088/1361-6382/abf89c}{Classical Quantum Gravity \textbf{38} 125003 (2021).}}
		
		\bibitem{BBCRG:22}  C.\ Barceló, V.\ Boyanov, R.\ Carballo-Rubio, and L.\ J.\ Garay,
		{\href{https://doi.org/10.1103/PhysRevD.106.124006}{Phys. Rev. D \textbf{106}, 124006 (2022).}}
		
		\bibitem{KS:16}  D. Kubizňák and F. Simovic,
		{\href{https://doi.org/10.1088/0264-9381/33/24/245001}{Classical Quantum Gravity \textbf{33}, 245001 (2016).}}
		
		\bibitem{SM:18}  F. Simovic and R. B Mann,
		{\href{https://doi.org/10.1088/1361-6382/aaf445}{Classical Quantum Gravity \textbf{36}, 014002 (2018).}}
		
		\bibitem{SM:19}  F. Simovic and R. B Mann,
		{\href{https://doi.org/10.1007/JHEP05(2019)136}{ J. High Energy Phys. 05 (2019) 136. }}
		
		\bibitem{HHMS:20}  S. Haroon, R. A. Hennigar, R. B. Mann, and F. Simovic,
		{\href{https://doi.org/10.1103/PhysRevD.101.084051}{Phys. Rev. D \textbf{101}, 084051 (2020).}}
		
		\bibitem{SFM:21} F. Simovic, D. Fusco, and R. B. Mann,
		{\href{https://doi.org/10.1007/JHEP02(2021)219}{J. High Energy Phys. 02 (2021) 219.}}
		
		\bibitem{S-essay:23} F. Simovic,
		{\href{https://doi.org/10.1142/S0218271823420233}{Int.\ J.\ Mod.\ Phys.\ D \textbf{32}, 2342023 (2023).}}
		
		\bibitem{LHK:23} M. Liška, R. A. Hennigar, and D. Kubizňák,
		\href{https://doi.org/10.1007/JHEP11(2023)195}{J. High Energy Phys. 11 (2023) 195}.
		
		\bibitem{HLJV:21} K. Hajian, S. Liberati, M. M. Sheikh-Jabbari, and M. H. Vahidinia,
		\href{https://doi.org/10.1016/j.physletb.2020.136002}{Phys. Let. B \textbf{812}, 136002 (2021)}.
		
		\bibitem{NN:23} S. Nojiri and G. G. L. Nashed, 
		{\href{https://doi.org/10.1103/PhysRevD.108.024014}{Phys. Rev. D \textbf{108}, 024014 (2023).}}
		
		\bibitem{GS:17} L. Gulin and I. Smolić,
		\href{https://doi.org/10.1088/1361-6382/aa9dfd}{Classical Quantum Gravity \textbf{35}, 025015 (2017)}.
		
		 \bibitem{landau} L. D. Landau and E. M. Lifshitz,
		\href{https://doi.org/10.1016/C2009-0-24487-4}{\textit{Statistical Physics} (Pergamon Press, Oxford, 1969)}.
		
		\bibitem{MS:23} S. Murk and I. Soranidis, 
		{\href{https://doi.org/10.1103/PhysRevD.108.124007}{Phys. Rev. D \textbf{108}, 124007 (2023).}}		
		
		\bibitem{ADM:59} R. Arnowitt, S. Deser, and C. W. Misner,
		{\href{https://doi.org/10.1103/PhysRev.116.1322}{Phys. Rev. \textbf{116}, 1322  (1959)}}.
		
		\bibitem{planck} Planck Collaboration,
		{\href{https://doi.org/10.1051/0004-6361/201833910}{Astron. Astrophys. \textbf{641}, A6 (2020)}}.
	
		
		\bibitem{FN:98} V. Frolov and I. Novikov,
		\href{https://doi.org/10.1007/978-94-011-5139-9}{\textit{Black Hole Physics, Basic Concepts and New Developments} (Springer Verlag, Heidelberg, 1998)}.
		
		\bibitem{S:95} S. N. Solodukhin,
		\href{https://doi.org/10.1103/PhysRevD.51.609}{Phys. Rev. D \textbf{51}, 609 (1995).}	    
	 		
		
	\end{thebibliography}
\end{document}